\documentclass[preprint,aps,prb]{revtex4-1}

\usepackage{amssymb}
\usepackage{amsbsy}
\usepackage{amsmath}
\usepackage{epsfig}
\usepackage{graphicx}

\begin{document}

\title{Graphene: Junctions and STM Spectra}

\author{M. Maiti}

\affiliation{The Institute of Mathematical Sciences,C.I.T Campus Taramani, Chennai-600113, India.}

\author{K. Saha}

\affiliation{Theoretical Physics Department, Indian Association for the Cultivation of Science, Kolkata-700032, India.}

\author{K. Sengupta}

\affiliation{Theoretical Physics Department, Indian Association for the Cultivation of Science, Kolkata-700032, India.}

\maketitle

\section{Introduction}\label{ra_sec1}

Graphene, a two-dimensional single layer of graphite, was first 
fabricated in 2004 by Novoselov {\it et.\,al.} \cite{nov1}. This
has provided an unique opportunity for experimental observation of
electronic properties of graphene which has attracted theoretical
attention for several decades \cite{oldref}. 
The importance of graphene lies not only in providing the first 
realization of Dirac physics in condensed matter systems but also 
in providing a way of realization of several devices in nanometer scale.  
In this article, we are going to concern ourselves mainly on the 
first of these two aspects of graphene.

In graphene, the energy bands touch the Fermi energy at six discrete 
points at the edges of the hexagonal Brillouin zone. Out of these 
six Fermi points, only two are inequivalent; they are commonly 
referred to as $K$ and $K'$ points \cite{ando1}. 
The quasiparticle excitations about these $K$
and $K'$ points obey linear Dirac-like energy dispersion. The
presence of such Dirac-like quasiparticles is expected to lead to a
number of unusual electronic properties in graphene including
relativistic quantum Hall effect with unusual structure of Hall
plateaus \cite{shar1}. Recently, experimental observation of the
unusual plateau structure of the Hall conductivity has confirmed
this theoretical prediction \cite{nov2}. Further, as suggested in
Ref.\ \onlinecite{geim1}, the presence of such quasiparticles in
graphene provides us with an experimental test bed for Klein
paradox. \cite{klein1} These and several other properties of
graphene has been covered extensively in several review articles
\cite{netorev1,rev2,rev3}. In the current article, we are going to
focus on the effect of the Dirac nature of graphene quasiparticles
on two separate aspects. The first of these involves transport of
superconducting graphene junctions while the second involves Kondo
effect and scanning tunneling spectra of graphene.

It is well known that the existence Dirac-like quasiparticles
affects tunneling conductance of a normal metal-superconductor (NS)
interface of graphene \cite{beenakker1}. Graphene is not a natural
superconductor. However, superconductivity can be induced in a
graphene layer in the presence of a superconducting electrode near
it via proximity effect \cite{volkov1,beenakker1,beenakker2} or by
possible intercalation with dopant molecules \cite{uchoa1}. It has
been recently predicted \cite{beenakker1} that a graphene NS
junction, due to the Dirac-like energy spectrum of its
quasiparticles, can exhibit specular Andreev reflection in contrast
to the usual retro reflection observed in conventional NS junctions
\cite{andreev1,tinkham1}. Such specular Andreev reflection process
leads to qualitatively different tunneling conductance curves
compared to conventional NS junctions \cite{beenakker1}. The effect
of the presence of a thin barrier region of thickness $d \rightarrow
0$ created by applying a large gate voltage $ V_0 \rightarrow
\infty$ (such that $V_0 d$ is finite) between the normal and the
superconducting region has also been studied in Ref.\
\onlinecite{sengupta1}. It has been shown that in this thin
barrier limit, in contrast to all normal
metal-barrier-superconductor (NBS) junctions studied so far, the
tunneling conductance of a graphene NBS junction is an oscillatory
function of the dimensionless barrier strength $\chi = V_0 d /(\hbar
v_F)$, where $v_F$ denotes the Fermi velocity of graphene, with
periodicity $\pi$. Further, it has also been demonstrated that the
tunneling conductance reaches its maxima of $2G_0$ for $ \chi =
(n+1/2)\pi$, where $n$ is an integer. The latter result was also
interpreted in terms of transmission resonance property of the
Dirac-Bogoliubov quasiparticles \cite{nov2}. However, no such
studies have been undertaken for NBS junctions with barriers of
arbitrary thickness $d$ and barrier potential $V_0$. As we shall
discuss in details in Sec.\ \ref{NBS}, the analysis of Ref.\ \onlinecite{sengupta1}
and calculate the tunneling conductance of a graphene NBS junction
with a barrier of thickness $d$ and with an arbitrary voltage $V_0$
applied across the barrier region can also be extended to thick
barrier junctions \cite{sengupta2jc}. It can be shown that the
oscillatory behavior of the tunneling conductance is not a property
of the thin barrier limit, but persists for arbitrary barrier width
$d$ and applied gate voltage $V_0$, as long as $d \ll \xi$, where
$\xi$ is the coherence length of the superconductor. Further, the
periodicity and amplitude of these oscillations deviate from their
values in the thin barrier limit and becomes a function of the
applied voltage $V_0$.

The study of Josephson effect in graphene for tunnel SBS junctions also
presents some unconventional features due to the presence of the
Dirac quasiparticles. In this review, we shall concentrate on SBS
junctions with barrier thickness $d \ll \xi$ where $\xi$ is the
superconducting coherence length, and width $L$ which has an applied
gate voltage $V_0$ across the barrier region \cite{maiti1}. The central property
of such junctions on which we shall mainly focus on is that in
complete contrast to the conventional Josephson tunnel junctions
studied so far \cite{likharev1jc,golubov1jc}, the Josephson current in
graphene SBS tunnel junctions is an oscillatory function of both the
barrier thickness $d$ and the applied gate voltage $V_0$. In the
thin barrier limit, where the barrier region can be characterized by
an effective dimensionless barrier strength $\chi = V_0 d/\hbar v_F$
($v_F$ being the Fermi velocity of electrons in graphene), the
Josephson current becomes an oscillatory function of $\chi$ with
period $\pi$ \cite{maiti1}. In this limit, the oscillatory
behavior of Josephson current can be understood as a consequence of
transmission resonance phenomenon of Dirac-Bogoliubov-de Gennes
(DBdG) quasiparticles in graphene. The Josephson current reaches the
Kulik-Omelyanchuk limit \cite{ko1jc} for $\chi=n \pi$ ($n$ being an
integer), but, unlike conventional junctions, never reaches the
Ambegaokar Baratoff limit \cite{ambe1jc} for large $\chi$. This analysis
is done in Sec.\ \ref{SBS}.

Another extremely interesting phenomenon in conventional metal
systems is the Kondo effect which occurs in the presence of dilute
concentration of localized quantum spins coupled to the
spin-degenerate Fermi sea of metal electrons \cite{kondoref1}. The
impurity spin-electron interaction then results in perfect or
partial screening of the impurity spin as one approaches zero
temperature. It also results in a sharp `Kondo Resonance' in
electron spectral functions. Recent developments in quantum dots and
nano devices have given new ways in which various theoretical
results in Kondo physics, which are not easily testable otherwise,
can be tested and confirmed experimentally \cite{kondoref2}. Most of
the early studies in Kondo effect were carried on for conventional
metallic systems with constant density of states (DOS) at the Fermi
surface \cite{affleck1}. Some studies on Kondo effect in possible
flux phases \cite{frad1}, nodal quasiparticles in d-wave
superconductors \cite{subir1}, Luttinger liquids \cite{furu1}, and
hexagonal Kondo lattice \cite{saremi}, for which the DOS of the
associated Fermions vanishes as some power law at the Fermi surface,
has also been undertaken. Recently, there has been a interest in
study of the physics of magnetic impurities in graphene.
\cite{baskaran1,martina1,saha,uchao2,hari1}  One of the purpose of
this article is to articulate a part of this recent progress in
Sec.\ \ref{Kondosec}.

Scanning tunneling microscopes (STM) are extremely useful probes for
studying properties of two or quasi-two dimensional materials
\cite{stmgra1,davis1}. Studying electronic properties of a sample
with STM typically involves measurement of the tunneling conductance
$G(V)$ for a given applied voltage $V$. The tunneling conductances
measured in these experiments have also been studied theoretically
for conventional metallic systems and are known to exhibit Fano
resonances at zero bias voltage in the presence of impurities
\cite{madhavan1,wingreen1}. The application and utility of this
experimental technique, with superconducting STM tips, has also been
discussed in the literature for conventional systems \cite{stmsc}.
However, tunneling spectroscopy of graphene using superconducting
STM tips remains to be studied both experimentally and
theoretically. In Sec.\ \ref{stmsection}, we shall elaborate the
progress on the STM response of doped graphene and discuss some of
it's unconventional features. For undoped graphene with Fermi energy
$E_F=0$, the derivative of the STM tunneling conductance ($G$) with
respect to the applied voltage ($dG/dV$) reflects the density of
states (DOS) of the STM tip ($\rho_t$), ${\it i.e.}$, $dG/dV \sim
+(-) \rho_t$ for $V>(<)0$. By tuning $E_F$, one can interpolate
between this unconventional $\rho_t \sim \pm dG/dV$ and the
conventional $\rho_t \sim G$ (seen for $E_F \gg eV$) behaviors.
Further, for superconducting STM tips with energy gap $\Delta_0$,
$G\, (dG/dV)$ displays a cusp (discontinuity) at $eV=-E_F-\Delta_0$
as a signature of the Dirac point which should be experimentally
observable in graphene with small $E_F$ where the regime $eV > E_F$
can be easily accessed. For impurity doped graphene with large
$E_F$, experiments in Ref.\ \onlinecite{hari1} have seen that the
tunneling conductance, as measured by a metallic STM tip, depends
qualitatively on the position of the impurity in the graphene
matrix. For impurity atoms atop the hexagon center, the zero-bias
tunneling conductance shows a peak; for those atop a graphene site,
it shows a dip. We provide a detailed discussion of this phenomenon
and point out that its origin lies in conservation/breaking of
pseudospin symmetry of the Dirac quasiparticles by the impurity.

The organization of the rest of the review is as follows.
We give a generic description of the graphene NBS and 
SBS tunnel junctions which is described by the Dirac-Bogoliubov-de Gennes
(DBdG) equations.
In section \ref{NBS} we review the theory of tunneling conductance
of a graphene NBS junction with a barrier of thickness $d$
and with an arbitrary voltage $V_0$ applied across the barrier region.
The results obtained are then compared and contrasted with that of a thin
barrier and zero barrier junction. In section
\ref{SBS} we study Josephson current for a general SBS junction
barrier region of thickness $d$ and potential $V_0$. We also discuss the thin 
barrier limit to understand the oscillatory behavior 
 in terms of transmission resonance of DBdG particles. 
Finally we study some possible experimental realizations of 
the above mentioned junctions to probe the 
oscillations in section \ref{experiments}.
In section. \ref{Kondosec}, we discuss the unconventional Kondo 
effect in graphene. We describe the large N analysis for a 
generic spin S local moment coupled to Dirac electrons 
in graphene. The analysis gives rise to a 
finite critical Kondo coupling strength which can be tuned
by the application of an external gate voltage 
and is particular to graphene. We also discuss the 
possible realization of the non-Fermi liquid ground 
states via the multichannel Kondo effect. 
In  section \ref{stmsection}, we discuss the STM response of 
graphene. We discuss the tunneling current through the STM tip
within linear-response theory using a superconding tip to probe an 
undoped sample and a metallic tip with constant density of 
states (DOS) to probe an impurity present in the sample.
We conclude with a general discussion on the unconventional 
tunneling, STM properties and the behaviour of magnetic impurities 
in graphene in section \ref{conclusion}.  

\section {Transport Properties of Superconducting Junctions}
\label{junctions}

\begin{figure}[htp]
\centerline{\epsfig{file=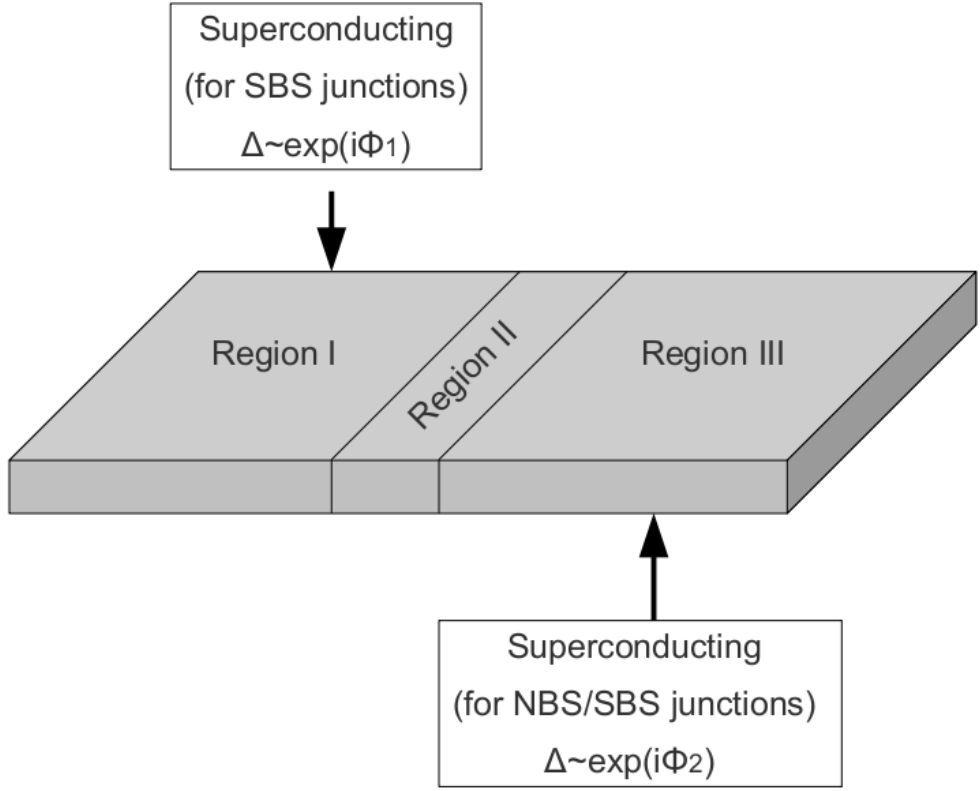, width=.9\linewidth}}
\caption{A schematic sketch of the graphene superconducting junction.
Region I denotes the normal(superconducting) region for NBS(SBS) junction. A potential
barrier $V_0$ of width $d$ is created with the application of the external gate voltage.
The region III is the superconducting region for both the NBS and SBS junctions. Superconductivity, as
 discussed in the text is induced in the grpahene layer by proximity effect. } \label{junction_fig}
\end{figure}

An understanding of the transport properties across different
superconducting junctions of graphene throws substantial light on the electronic
properties. 
A generic description of the junctions to study the transport properties 
is as follows. A local potential
barrier of width $d$ is implemented on the
graphene sheet occupying the $xy$ plane 
by either using the electric field effect or 
local chemical doping \cite{nov2, zhang1, geim1}
A $s$-wave pairing $\Delta(\mathbf{r})$ is induced in graphene 
via proximity effect \cite{beenakker1,heersche1}.
For NBS (SBS) region \textbf{I} is normal (superconducting)
region occupying $x \leq −d$ for all y as shown schematically in
Fig. \ref{junction_fig}. The
region \textbf{II} modeled by a barrier potential $V_0$, 
extends from $x = −d$ to $x = 0$
while the superconducting region occupies x $\geq$ 0 
(marked as region \textbf{III} in Fig (\ref{junction_fig}).
For calculations we shall assume
that the barrier region has sharp edges on both sides. This condition requires that
$d \ll \lambda = 2\pi/k_F$ , where $k_F$ and $\lambda$ are
Fermi wave-vector and wavelength for graphene, and can be
realistically created in experiments \cite{geim1}.
Also, the interface is smooth and impurity free on the
scale of the superconducting coherence length $\xi = \hbar v_F/\Delta_0$,
where $\Delta_0$ is the amplitude of the induced superconducting order
parameter. For both the junctions (NBS and SBS) the induced pair potential
can be modeled (with appropriate boundary condition for the two different junctions) as:
\begin{equation}
\Delta({\bf r}) = \Delta_0 \exp(i\phi) \label{deltaeq}
\end{equation}
$\phi$ is the phase. These junctions can then be described
by the Dirac-Bogoliubov-de Gennes (DBdG) equations: 
\begin{eqnarray}
&&\left(\begin{array}{cc}
    {\mathcal H}_{a}-E_F + U({\bf r}) & \Delta ({\bf r}) \\
     \Delta^{\ast}({\bf r}) & E_F - U({\bf r})-{\mathcal H}_{a}
    \end{array}\right) \psi_{a}   = E \psi_{a}
\label{bdg1}
\end{eqnarray}
Here, $\psi_a = \left(\psi_{A\,a}, \psi_{B\,a}, \psi_{A\,{\bar
a}}^{\ast}, -\psi_{B\,{\bar a}}^{\ast}\right)$ are the $4$ component
wavefunctions for the electron and hole spinors, the index $a$
denote $K$ or $K'$ for electron/holes near $K$ and $K'$ points,
${\bar a}$ takes values $K'(K)$ for $a=K(K')$, $E_F$ denote the
Fermi energy. $A$ and $B$ denote the
two inequivalent sites in the hexagonal lattice of graphene, and the
Hamiltonian ${\mathcal H}_a$ is given by
\begin{eqnarray}
{\mathcal H}_a &=& -i \hbar v_F \left(\sigma_x \partial_x + {\rm
sgn}(a) \sigma_y
\partial_y \right). \label{bdg2}
\end{eqnarray}
In Eq.\ \ref{bdg2}, $v_F$ denotes the Fermi velocity of the
quasiparticles in graphene and ${\rm sgn}(a)$ takes values $\pm$ for
$a=K(K')$. The potential $U({\bf r})$ gives the relative shift of Fermi energies in
the barrier and superconducting regions and is modeled as:
\begin{eqnarray}
U({\bf r}) = V_0 \theta(-x) \theta(x+d) \label{poteqjc}
\end{eqnarray}
Eq.\ \ref{bdg1} can be solved in a straightforward manner to yield
the wavefunction $\psi$ in the normal, insulating and the
superconducting regions, taking into account both Andreev and normal reflection
processes. These wavefunctions satisfy the
appropriate boundary conditions at the interfaces of the junctions.
Note however that these boundary conditions, in contrast their
counterparts in standard junction interfaces, do not impose any
constraint on derivative of the wavefunctions at the boundary.
The tunneling conductance and Josephson current across the junctions
can then be calculated using appropriate expressions. These are
found to have novel oscillatory behavior in complete contrast to their
standard counterparts as will be described in the subsequent sections.

\subsection{NBS junction} \label{NBS}

The pair-potential for the NBS junction is modeled as:
\begin{eqnarray}
\Delta({\bf r}) = \Delta_0 \exp(i\phi) \theta(x), \label{deltaeqnb}
\end{eqnarray}
$\theta(x)$ is the Heaviside step function.
Eq.\ \ref{bdg1}  can be solved in a straightforward manner to yield
the wavefunction $\psi$ in the normal, insulating and the
superconducting regions. In the normal region, for electron and
holes traveling the $\pm x$ direction with a transverse momentum
$k_y=q$ and energy $\epsilon$, the (unnormalized) wavefunctions
are given by
\begin{eqnarray}
\psi_N^{e \pm} &=&  \left(1,\pm e^{\pm i \alpha},0,0\right) \exp
\left[i \left(\pm k_{n} x + q y\right) \right], \nonumber\\
\psi_N^{h \pm} &=& \left(0,0,1,\mp e^{\pm i \alpha'}\right) \exp
\left[i \left(\pm k'_{n} x + q y \right)\right],
\nonumber\\
\sin(\alpha) &=& \frac{\hbar v_F q}{\epsilon +E_F},  \quad
\sin(\alpha') = \frac{\hbar v_F q}{\epsilon - E_F}, \label{wavenorm}
\end{eqnarray}
where the wave-vector $k_n (k'_n)$ for the electron (hole)
wavefunctions are given by
\begin{eqnarray}
k_n(k'_n) &=& \sqrt{\left(\frac{\epsilon +(-) E_F}{\hbar
v_F}\right)^2-q^2 }, \label{ehwave}
\end{eqnarray}
and $\alpha (\alpha')$ is the angle of incidence of the electron
(hole). 

In the barrier region, one can similarly obtain
\begin{eqnarray}
\psi_B^{e \pm} &=& \left(1,\pm e^{\pm i \theta},0,0\right) \exp
\left[i\left(\pm k_{b} x + q y \right)\right], \nonumber\\
\psi_B^{h \pm} &=& \left(0,0,1,\mp e^{\pm i \theta'}\right) \exp
\left[i \left(\pm k'_{b} x + q y \right)\right], \label{barrwave}
\end{eqnarray}
for electron and holes moving along $\pm x$. Here the angle of
incidence of the electron(hole) $\theta(\theta')$ and the wavevector
$k_b(k'_b)$ are given by is
\begin{eqnarray}
\sin\left[\theta(\theta')\right] &=& \hbar v_F q/\left[\epsilon
+(-)(E_F-V_0)\right], \nonumber\\
k_b(k'_b) &=& \sqrt{\left(\frac{\epsilon +(-) (E_F-V_0)}{\hbar
v_F}\right)^2 -q^2}. \label{barrwave2}
\end{eqnarray}
Note that Eq.\ \ref{barrwave} ceases to be the solution of the Dirac
equation (Eq.\ \ref{bdg1}) when $E_F=V_0$ and $\epsilon=0$. For
these parameter values, Eq.\ \ref{bdg1} in the barrier region
becomes $ {\mathcal H}_a \psi_B =0$ which do not have purely
oscillatory solutions. For the rest of the calculation, we shall restrict
ourselves to the regime $V_0 > E_F$.

In the superconducting region, the DBdG quasiparticles are mixtures
of electron and holes. Consequently, the wavefunctions of the DBdG
quasiparticles moving along $\pm x$ with transverse momenta $q$ and
energy $\epsilon$, for $(U_0+E_F) \gg \Delta_0, \epsilon$, has the
form
\begin{eqnarray}
\psi_{I}^{\pm} &=& \left( w_1^{\pm}, w_2^{\pm},w_3^{\pm},w_4^{\pm}
\right) e^{\left[ i\left(\pm k_s x +q y\right) + \kappa x\right]}
 \label{supwave}
\end{eqnarray}
where
\begin{eqnarray}
\frac{w_2^{\pm}}{w_1^{\pm}} &=& \pm \exp(\pm i\gamma), \quad
\frac{w_3^{\pm}}{w_1^{\pm}} = \exp[-i(\phi_1 \mp \beta)],\nonumber\\
\frac{w_4^{\pm}}{w_1^{\pm}} &=& \pm \exp[\pm i(\mp \phi_1 +\beta +
\gamma)], \label{ratieq}
\end{eqnarray}
where $\gamma$ is the angle of incidence for the quasiparticles.
Here the wavevector $k_s$ and the localization length $\kappa^{-1}$
can be expressed as a function of the energy $\epsilon$ and the
transverse momenta $q$ as
\begin{eqnarray}
k_s &=& \sqrt{\left[\left(U_0+E_F\right)/\hbar v_F\right]^2 -q^2},
\nonumber\\
\kappa^{-1} &=&  \frac{(\hbar v_F)^2 k_s}{\left[(U_0+E_F) \Delta_0
\sin(\beta)\right]}, \label{local}
\end{eqnarray}
where $\beta$ is given by
\begin{eqnarray}
\beta &=& \cos^{-1} \left(\epsilon/\Delta_0\right) \quad {\rm if}
\left|\epsilon\right| < \Delta_0 ,\nonumber\\
&=& -i \cosh^{-1} \left(\epsilon/\Delta_0\right) \quad {\rm if}
\left|\epsilon\right| > \Delta_0.\label{betaeq}
\end{eqnarray}
Note that for $\left|\epsilon\right| > \Delta_0$, $\kappa$ becomes
imaginary and the quasiparticles can propagate in the bulk of the
superconductor.

Next we note that for the Andreev process to take place, the angles
$\theta$, $\theta'$ and $\alpha'$ must all be less than
$90^{\circ}$. This sets the limit of maximum angle of incidence
$\alpha$. Using Eqns.\ \ref{wavenorm} and \ref{barrwave2}, one finds
that the critical angle of incidence is
\begin{eqnarray}
\alpha_c &=& \alpha_c^{(1)} \theta(V_0-2E_F) + \alpha_c^{(2)}
\theta(2E_F-V_0)  \nonumber\\
\alpha_c^{(1)} &=&
\arcsin\left[\left|\epsilon -
E_F\right|/\left(\epsilon + E_F\right)\right], \nonumber\\
\alpha_c^{(2)} &=& \arcsin\left[\left|\epsilon -
|E_F-V_0|\right|/\left(\epsilon + E_F\right)\right]. \label{criti1}
\end{eqnarray}
Note that in the thin or zero barrier limits treated in Refs.\
\onlinecite{sengupta1} and \onlinecite{beenakker1},
$\alpha_c=\alpha_c^{(1)}$ for all parameter regimes.

\begin{figure}
\centerline{\epsfig{file=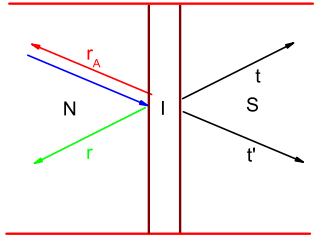, width=6cm}}
\caption{A schematic sketch of normal reflection
($r$), Andreev reflection ($r_A$) and transmission processes ($t$
and $t'$) at a graphene NBS junction. Note that in this schematic
picture, we have chosen $r_A$ to denote a retro Andreev reflection
for illustration purpose. In practice, as discussed in the text,
$r_A$ takes into account possibilities of both retro and specular
Andreev reflections. The electron and hole wavefunctions inside the
barrier region is not sketched to avoid clutter.} \label{figprocess}
\end{figure}

Let us now consider a electron-like quasiparticle incident on the
barrier from the normal side with an energy $\epsilon$ and
transverse momentum $q$. The basic process of ordinary and Andreev
reflection that can take place at the interface is schematically
sketched in Fig.\ \ref{figprocess}. As noted in Ref.\ \onlinecite{beenakker1},
in contrast to conventional NBS junction,
graphene junctions allow for both retro and specular Andreev
reflections. The former dominates when $\epsilon, \Delta_0 \ll E_F$
so that $ \alpha = -\alpha'$ (Eq.\ \ref{wavenorm}) while that latter
prevails when $E_F \ll \epsilon, \Delta_0$ with $\alpha = \alpha'$.
Note that in Fig.\ \ref{figprocess}, we have chosen $r_A$ to denote
a retro Andreev reflection for illustration purposes. In practice,
$r_A$ includes both retro and specular Andreev reflections. In what
follows, we shall denote the total probability amplitude of Andreev
reflection as $r_A$ which takes into account possibilities of both
retro and specular Andreev reflections.

The wave functions in the normal, insulating and superconducting
regions, taking into account both Andreev and normal reflection
processes, can then be written as \cite{tinkham1}
\begin{eqnarray}
\Psi_N &=& \psi_N^{e +}+r \psi_N^{e -} + r_A \psi_N^{h -}, \quad
\Psi_S = t \psi_S^{+}+ t' \psi_S^{-}, \nonumber\\
\Psi_B &=& p \psi_B^{e +}+q \psi_B^{e -} + m \psi_B^{h +} + n
\psi_N^{h -}, \label{wave2}
\end{eqnarray}
where $ r$ and $r_A$ are the amplitudes of normal and Andreev
reflections respectively, $t$ and $t'$ are the amplitudes of
electron-like and hole-like quasiparticles in the superconducting
region and $p$, $q$, $m$ and $n$ are the amplitudes of electron and
holes in the barrier. These wavefunctions must satisfy the
appropriate boundary conditions:
\begin{eqnarray}
\Psi_N |_{x=-d} &=& \Psi_B |_{x=-d},  \quad  \Psi_B |_{x=0} = \Psi_S
|_{x=0}. \label{bc1}
\end{eqnarray}
These boundary conditions yield eight linear homogeneous equations for the
coefficients $r$, $r_A$, $t$, $t^{'}$, $p$, $q$, $m$, and $n$.

After some straightforward but cumbersome algebra, we find that
\begin{eqnarray}
r &=& e^{-2i k_n d} \frac{\mathcal N}{\mathcal D}, \label{req} \\
{\mathcal N} &=& \left[e^{i\alpha}\cos(k_{b}d+\theta)-i \sin(k_{b}d)\right]\nonumber\\
&& -\rho[\cos(k_{b}d-\theta)-i\ e^{i\alpha} \sin(k_{b}d)], \label{neq}\\
{\mathcal D} &=& \left [e^{-i\alpha}\cos(k_{b}d+\theta)
+i\sin(k_{b}d)\right]\nonumber\\
&& + \rho\left[\cos(k_{b}d-\theta)+ie^{-i\alpha}\sin(k_{b}d)\right],  \label{deq} \\
t'&=&{\frac{e^{-ik_n d}}{\cos(\theta) [\Gamma
e^{-i\beta}+e^{i\beta}]}} \Big([\cos(k_{b}d-\theta)-ie^{i\alpha} \sin(k_{b}d)]\nonumber\\
&& +r e^{ik_n d}[\cos(k_{b}d-\theta)+ie^{-i\alpha}\sin(k_{b}d)] \Big),\\
t &=& \Gamma t',\\
r_A &=& \frac{ t (\Gamma + 1)e^{ik'_n d}
\cos(\theta')e^{-i\phi}}{\cos(k'_{b}d-\theta')-ie^{-i\alpha'}
\sin(k'_{b}d)}, \label{raeq}
\end{eqnarray}
where the parameters $\Gamma$ and $\rho$ can be expressed in terms
of $\gamma$, $\beta$, $\theta$, $\theta'$, $\alpha$, and $\alpha'$
(Eqs.\ \ref{wavenorm}, \ref{barrwave2}, \ref{supwave}, and
\ref{betaeq}) as
\begin{eqnarray}
\rho &=& \frac{-\Gamma e^{i(\gamma-\beta)}+e^{-i(\gamma-\beta)}}
{\Gamma e^{-i\beta}+e^{i\beta}},\\
\Gamma &=& \frac{e^{-i\gamma}-\eta}{e^{i\gamma}+\eta},\\
\eta &=& \frac{e^{-i\alpha'} \cos(k'_{b}d+\theta')-i\sin(k'_{b}d)}
{\cos(k'_{b}d-\theta')-ie^{-i\alpha'} \sin(k'_{b}d)}. \label{qt1}
\end{eqnarray}
The tunneling conductance of the NBS junction can now be expressed
in terms of $r$ and $r_A$ by \cite{tinkham1}
\begin{eqnarray}
\frac{G(eV)}{G_0(eV)} &=&  \int_0^{\alpha_c}
\left(1-\left|r\right|^2 + \left|r_A\right|^2
\frac{\cos(\alpha')}{\cos(\alpha)} \right) \cos(\alpha) \, d\alpha,
\nonumber \\ \label{tc1}
\end{eqnarray}
where $G_0 = 4e^2 N(eV)/h$ is the ballistic conductance of metallic
graphene, $eV$ denotes the bias voltage, and $N(\epsilon)= (E_F
+\epsilon)w/(\pi \hbar v_F)$ denotes the number of available
channels for a graphene sample of width $w$. For $eV \ll E_F$, $G_0$
is a constant. Eq.\ \ref{tc1} can be evaluated numerically to yield
the tunneling conductance of the NBS junction for arbitrary
parameter values. We note at the outset, that $G=0$ when
$\alpha_c=0$. This occurs in two situations. First, when $eV=E_F$
and $V_0 \ge 2E_F$ so that $\alpha_c=\alpha_c^{(1)}$ vanishes. For
this situation to arise, $E_F +U_0> \Delta > E_F$ which means that
$U_0$ has to be finite. Second, $\alpha_c=\alpha_c^{(2)}=0$ when
$eV=0$ and $E_F=V_0$, so that the zero-bias conductance vanishes
when the barrier potential matches the Fermi energy of the normal
side 
\cite{comment1}
%

We now make contact with the results of
the thin barrier limit. We note that
since there are no condition on the derivatives of wavefunctions in
graphene NBS junctions, the standard delta function potential
approximation for thin barrier \cite{tinkham1} can not be taken the
outset, but has to be taken at the end of the calculation. This
limit is defined as $d/\lambda \rightarrow 0$ and $V_0/E_F
\rightarrow \infty$ such that the dimensionless barrier strength
\begin{eqnarray}
\chi &=&  V_0 d/\hbar v_F = 2\pi \left(\frac{V_0}{E_F}\right) \left(
\frac{d}{\lambda}\right) \label{barstr}
\end{eqnarray}
remains finite. In this limit, as can be seen from Eqs.\
\ref{wavenorm}, \ref{barrwave2} and \ref{supwave}, $\theta, \theta',
k_n d, k'_n d \rightarrow 0$ and $k_b d, k'_b d \rightarrow \chi$ so
that the expressions for $\Gamma$, $\rho$ and $\eta$ (Eq.\
\ref{qt1})
\begin{eqnarray}
\Gamma^{\rm tb} &=& \frac{e^{-i\gamma} -\eta^{\rm tb}}{e^{i\gamma}
+\eta^{\rm tb}}, \quad \eta^{\rm tb}
 = \frac{e^{- i \alpha'} \cos(\chi)  - i \sin(\chi)}{
\cos(\chi) - i e^{-i \alpha'} \sin(\chi)}, \nonumber\\
\rho^{\rm tb} &=& \frac{e^{-i(\gamma - \beta)} - \Gamma^{\rm tb}
e^{i(\gamma - \beta)}}{\Gamma^{\rm tb} e^{-i\beta} + e^{i \beta}}.
\label{coefftb1}
\end{eqnarray}
where the superscript "${\rm tb}$" denotes thin barrier. Using the
above-mentioned relations, we also obtain
\begin{eqnarray}
r^{\rm tb} &=& \frac{\cos(\chi) \left(e^{i \alpha}-\rho^{\rm tb}
\right) - i \sin(\chi)\left(1-\rho^{\rm tb} e^{i \alpha}
\right)}{\cos(\chi) \left(e^{-i \alpha}+\rho^{\rm tb}\right) + i
\sin(\chi)\left(1+\rho^{\rm tb} e^{-i \alpha}
\right)},\nonumber\\
 t^{'{\rm tb}} &=& \frac{\cos(\chi) \left(1+r^{\rm tb}\right) - i \sin(\chi)\left(e^{i
\alpha}-r^{\rm tb} e^{-i \alpha}\right)}{\Gamma e^{-i\beta} + e^{i
\beta}}, \nonumber\\
t^{\rm tb} &=&  \Gamma t^{'{\rm tb}},\nonumber\\
r_A^{\rm tb} &=& \frac{t'^{\rm tb} \left(\Gamma+1\right) e^{-i
\phi}}{\cos(\chi) - i e^{-i \alpha'} \sin(\chi)}. \label{coefftb2}
\end{eqnarray}
Eqs. \ref{coefftb1} and \ref{coefftb2} are precisely the result
obtained in Ref.\ \onlinecite{sengupta1} 
for the tunneling conductance of a thin graphene NBS junction.
The result obtained for a zero barrier in Ref.\ \onlinecite{beenakker1}
can be now easily obtained from Eqs.\
\ref{coefftb1} and \ref{coefftb2} by substituting $\chi=0$ in these
equations.

\subsubsection {Qualitative Discussions}

In this section, we shall analyze the formulae for tunneling
conductance obtained in above section. First we aim to obtain a
qualitative understanding of the behavior of the tunneling
conductance for finite barrier strength. To this end, we note from
Eq.\ \ref{tc1} that the maxima of the tunneling conductance must
occur where $|r|^2$ is minimum. In fact, if $|r|^2=0$ for all
transverse momenta, the tunneling conductance reaches its value
$2G_0$. Therefore we shall first try to analyze the expression of
$r$ (Eq.\ \ref{req}) for subgap voltages and when the Fermi surfaces
of the normal and superconducting sides are aligned with each other
($U_0=0$). In this case, we need $\Delta_0 \ll E_F$. So for subgap
tunneling conductance, we have $\epsilon \le \Delta_0 \ll E_F$. In
this limit, $\alpha \simeq -\alpha' \simeq \gamma$ (Eqs.
\ref{wavenorm} and \ref{supwave}), $k_b \simeq k_b'$, and $\theta
\simeq -\theta'$ (Eq. \ref{barrwave2}). Using these, one can write
\begin{eqnarray}
\eta &=& \frac{e^{i\alpha} \cos(k_b d - \theta) - i\sin(k_b
d)}{\cos(k_b d + \theta) - i e^{i \alpha} \sin(k_b d)}, \\
\rho &=& \frac{ \eta \cos(\alpha -\beta) + i
\sin(\beta)}{\cos(\alpha + \beta) + i \eta \sin(\beta)}.
\label{rhoeta}
\end{eqnarray}
Substituting Eq.\ \ref{rhoeta} in the expression of ${\mathcal N}$,
we find that the numerator of the reflection amplitude $r$ becomes
(Eqs.\ \ref{req} and \ref{neq})
\begin{eqnarray}
{\mathcal N} &=& \frac{e^{i \alpha}}{D_0} \Bigg[ -4 \sin(\alpha)
\sin(\beta) \cos(k_b d -\theta) \nonumber\\
&& \times \Big[- i\cos(\alpha) \sin(k_b d)\nonumber\\
&& + (\cos(k_b d - \theta)+\cos(k_b d + \theta))/2 \Big] \nonumber\\
&& + 2 \left[\cos(k_b d + \theta)-\cos(k_b d - \theta) \right]
\nonumber\\
&& \times \Big [ \cos(\alpha-\beta) \left\{ \cos(\alpha) +
\left[\cos(k_b
d - \theta) \right. \right. \nonumber\\
&& \left. \left.+ \cos(k_b d + \theta) \right]/2 \right\}  +
\sin(k_B
d) \sin(\beta)\Big] \Bigg], \label{rexp} \\
 D_0 &=&  \cos(k_b d + \theta)\cos(\alpha +
\beta) + \sin(k_b d)
\sin(\beta) \nonumber\\
&& + i e^{i \alpha} \left[ \cos(k_b d - \theta) \sin(\beta) -
\sin(k_b d) \cos(\alpha + \beta) \right]. \nonumber\\ \label{denexp}
\end{eqnarray}

\begin{figure}
\rotatebox{-90}{
\centerline{\epsfig{file=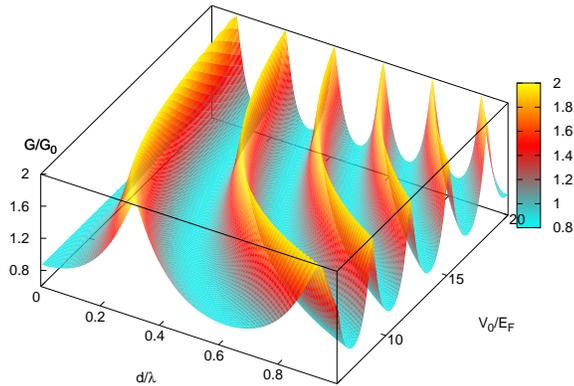, width=6cm}}}
\caption{Plot of zero-bias tunneling conductance for
$U_0=0$ and $\Delta_0=0.01 E_F$ as a function of gate voltage $V_0$
and barrier thickness $d$. Note that the oscillatory behavior of the
tunneling conductance persists for the entire range of $V_0$ and
$d$.} \label{figm1}
\end{figure}

From the expression of ${\mathcal N}$ (Eq.\ \ref{rexp}), we note the
following features. First, for normal incidence ($\alpha=0$) where $
\theta=\theta'=0$, ${\mathcal N}$ and hence $r$ (Eq.\ \ref{req})
vanishes. Thus the barrier is reflectionless for quasiparticles
which incident normally on the barrier for arbitrary barrier
thickness $d$ and strength of the applied voltage $V_0$. This is a
manifestation of Klein paradox for Dirac-Bogoliubov quasiparticles
\cite{klein1}. However, this feature is not manifested in tunneling
conductance $G$ ( Eq.\ \ref{tc1}) which receives contribution from
all angles of incidence. Second, apart from the above-mentioned
cases, $r$ never vanishes for all angles of incidence $\alpha$ and
arbitrary $eV < \Delta_0$ unless $\theta= \theta'$. Thus the subgap
tunneling conductance is not expected to reach a maximum value of
$2G_0$ as long as the thin barrier limit is not satisfied. However,
in practice, for barriers with $V_0>4E_F$, the difference between
$\theta$ and $\theta'$ turns out to be small for all $q \le k_F$
($\le 0.25$ for $q\le k_F$ and $eV=0$) so that the contribution to
${\mathcal N}$ (Eq.\ \ref{rexp}) from the terms $\sim (\cos(k_b d +
\theta) -\cos(k_b d -\theta))$ becomes negligible. Thus $|r|^2$ can
become quite small for special values of $V_0$ for all $q \le k_F$
so that the maximum value of tunneling conductance can reach close
to $2G_0$. Third, for large $V_0$, for which the contribution of
terms $\sim (\cos(k_b d + \theta) -\cos(k_b d -\theta))$ becomes
negligible, ${\mathcal N}$ and hence $r$ becomes very small when the
applied voltage matches the gap edge ${\it i.e.}$ $\sin (\beta)=0$
(Eq.\ \ref{rexp}). Thus the tunneling conductance curves approaches
close to its maximum value $2G_0$ and becomes independent of the
gate voltage $V_0$ at the gap edge $eV=\Delta_0$ for $\Delta_0 \ll
E_F$, as is also seen for conventional NBS junctions
\cite{tinkham1}. Fourth, in the thin barrier limit, ($V_0/E_F
\rightarrow \infty$ and $d/\lambda \rightarrow 0$), $\theta
\rightarrow 0$ and $k_b d \rightarrow \chi$, so that the
contribution of the terms $\sim (\cos(k_b d + \theta) -\cos(k_b d
-\theta))$ in Eq.\ \ref{rexp} vanishes and one gets

\begin{figure}
\centerline{\epsfig{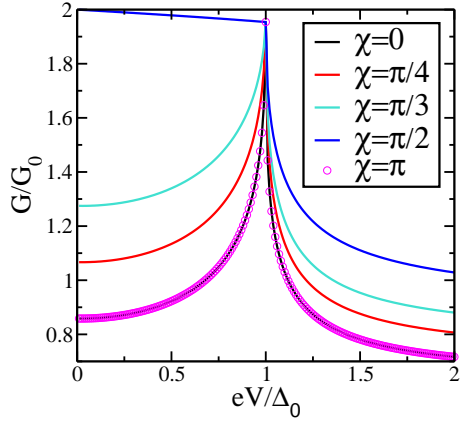}}
\caption{Plot of tunneling conductance of a NBS junction graphene as
a function of bias voltage for different effective barrier strengths
for $U_0=0$ and $\Delta_0=0.01 E_F$. Note that the curves for
$\chi=0$ (black line) and $\chi=\pi$(pink circles) coincide
reflecting $\pi$ periodicity.  } \label{fig1_tc_tb}
\end{figure}

\begin{eqnarray}
{\mathcal N}^{tb} &=& \frac{2 \sin(\alpha) [\sin(\chi + \beta)-
\sin(\chi-\beta)] }{D_0^{\rm tb} } \nonumber\\
&& \times \left[ -\cos(\chi) + i \sin(\chi) \cos(\alpha) \right],
\label{ntbeq} \\
D_0^{\rm tb} &=& \cos(\chi) \cos(\alpha + \beta) + \sin(\chi)
\sin(\beta) + i
e^{i \alpha} \nonumber\\
&& \times \left[ \cos(\chi) \sin(\beta) - \sin(\chi) \cos(\alpha +
\beta) \right].
\end{eqnarray}
${\mathcal N}^{tb}$
and hence $r^{\rm tb}$ (Eq.\ \ref{coefftb2}) vanishes at $\chi =
(n+1/2) \pi$ which yields the transmission resonance condition for
NBS junctions in graphene and is given in Fig.\ref{fig1_tc_tb} \cite{sengupta1}.

\begin{figure}
\centerline{\epsfig{file=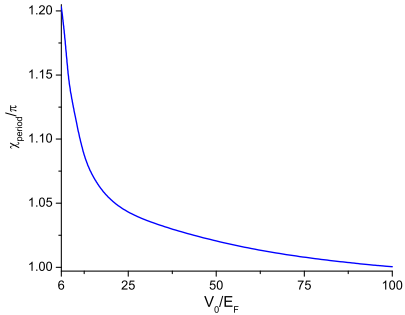, width=6cm}}
\caption{Plot of periodicity $\chi_{\rm period}$ of oscillations of
tunneling conductance as a function of applied gate voltage $V_0$
for $U_0=0$ and $\Delta_0=0.01 E_F$. Note that the periodicity
approaches $\pi$ as the voltage increases since the junction
approaches the thin barrier limit. } \label{figm2}
\end{figure}

Fifth, as can seen from Eqs.\ \ref{req}
and \ref{raeq}, both $|r|^2$ and $|r_A|^2$ are periodic functions of
$V_0$ and $d$ since both $k_b$ and $\theta$ depend on $V_0$. Thus
the oscillatory behavior of subgap tunneling conductance as a
function of applied gate voltage $V_0$ or barrier thickness $d$ is a
general feature of graphene NBS junctions with $d \ll \xi$. However,
unlike the thin barrier limit, for an arbitrary NBS junction, $k_b d
= \chi \sqrt{ (E_F/V_0-1)^2 + \hbar^2 v_F^2 q^2 /V_0^2} \neq \chi$,
and $\theta \neq 0$. Thus the period of oscillations of $|r|^2$ and
$|r_A|^2$ will depend on $V_0$ and should deviate from their
universal value $\pi$ in the thin barrier limits \cite{sengupta1}. Finally, we note
from Eqs.\ \ref{req}, \ref{tc1} and \ref{ntbeq} that in the thin
barrier limit (and therefore for large $V_0$), the amplitude of
oscillations of the zero-bias conductance for a fixed $V_0$, defined
as $[G_{\rm max}(eV=0;V_0)-G_{\rm min}(eV=0;V_0)]/G_0$, which
depends on the difference of $|r(\chi=(n+1/2)\pi)|^2$ and
$|r(\chi=n\pi)|^2$ becomes independent of $\chi$ or the applied gate
voltage $V_0$.

\begin{figure}
\centerline{\epsfig{file=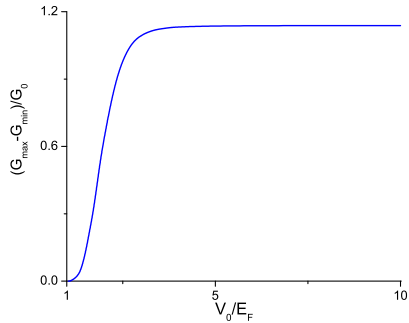,width=6cm}}
\caption{Plot of the amplitude $[G_{\rm max}(eV=0;V_0)-G_{\rm
min}(eV=0;V_0)]/G_0 \equiv (G_{\rm max}-G_{\rm min})/G_0$ of
zero-bias tunneling conductance as a function of the applied gate
voltage $V_0$ for $U_0=0$ and $\Delta_0=0.01 E_F$. Note that $G$
reaches $2G_0$ for $V_0 \ge 4E_F$ where the amplitude become
independent of the applied gate voltage as in the thin barrier limit
and vanishes for $V_0/E_F=1$ as discussed in the text.}
\label{figm3}
\end{figure}

\subsubsection{Numerical Results}

\begin{figure}
\rotatebox{-90}{
\centerline{\epsfig{file=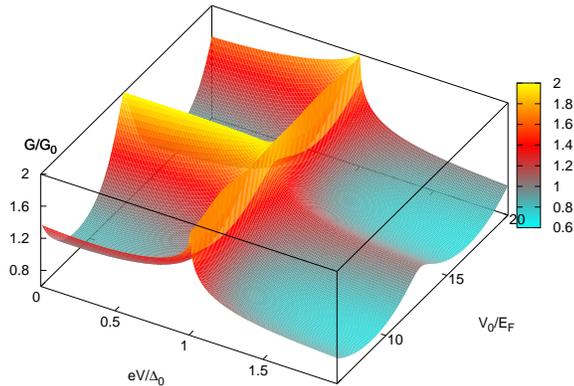, width=6cm}}}
\caption{Plot of tunneling conductance as a function
of the bias-voltage $eV$ and gate voltage $V_0$ for $d=0.4 \lambda$
and $\Delta_0=0.01 E_F$. Note that for large $V_0$, the tunneling
conductance at $eV= \Delta_0$ is close to $2G_0$ and becomes
independent of $V_0$ (see text for discussion).} \label{figm4}
\end{figure}

The above-mentioned discussion is corroborated by numerical
evaluation of the tunneling conductance as shown in Figs.\
\ref{figm1}, \ref{figm2}, \ref{figm3} and \ref{figm4}. From Fig.\
\ref{figm1}, which plots zero-bias tunneling conductance $G(eV=0)$
as a function of $V_0$ and $d$, we find that $G(eV=0)$ is an
oscillatory function of both $V_0$ and $d$ and reaches close to its
maximum value of $2G_0$ throughout the plotted range of $V_0$ and
$d$. Further, as seen from Fig.\ \ref{figm2}, the periodicity of
these oscillations becomes a function of $V_0$. To measure the
periodicity of these oscillations, the tunneling conductance is
plotted for a fixed $V_0$ as a function of $d$. The periodicity of
the conductance $d_{\rm period}$ is noted down from these plots and
$\chi_{\rm period} = V_0 d_{\rm period}/\hbar v_F$ is computed.
Fig.\ \ref{figm2} clearly shows that $\chi_{\rm period}$ deviate
significantly from their thin barrier value $\pi$ for low enough
$V_0$ and diverges at $V_0 \to E_F$ 
\cite{comment2}
. Fig.\
\ref{figm3} shows the amplitude of oscillations of zero-bias
conductance as a function of $V_0$. We note that maximum of the
zero-bias tunneling conductance $G_{\rm max}(eV=0)$ reaches close to
$2G_0$ for $V_0 \ge V_{0c} \simeq 4E_F$. For $V \ge V_{0c}$, the
amplitude becomes independent of the applied voltage as in the thin
barrier limit, as shown in Fig.\ \ref{figm3}. For $V_0 \to E_F$,
$\alpha_c=\alpha_c^{(2)} \to 0$, so that $G(eV=0) \to 0$ and hence
the amplitude vanishes. Finally, in Fig.\ \ref{figm4}, we plot the
tunneling conductance $G$ as a function of the applied bias-voltage
$eV$ and applied gate voltage $V_0$ for $d=0.4 \lambda$. We find
that, as expected from Eq.\ \ref{ntbeq}, $G$ reaches close to $2G_0$
at the gap edge for all $V_0 \ge 6E_F$. Also, as in the thin barrier
limit, the oscillation amplitudes for the subgap tunneling
conductance is maximum at zero-bias and shrinks to zero at the gap
edge $eV=\Delta_0$, where the tunneling conductance become
independent of the gate voltage.

\begin{figure}
\centerline{\epsfig{file=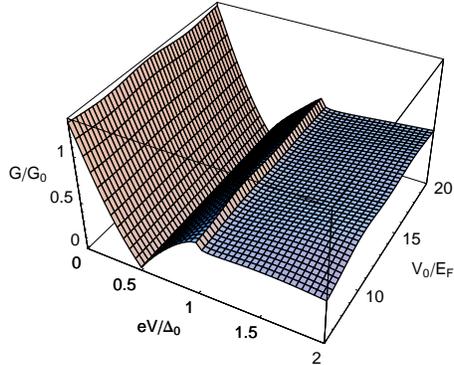, width=6cm}}
\caption{Plot of tunneling conductance as a function
of the bias-voltage $eV$ and the gate voltage $V_0$ for $d=0.4
\lambda$, $\Delta_0=2E_F$ and $U_0 =25 E_F$. As discussed in the
text, the tunneling conductance is virtually independent of the
applied gate voltage $V_0$ due to the presence of a large $U_0$.
Note that maximum angle of incidence for which Andreev reflection
can take place vanishes at $eV = E_F$ leading to vanishing of $G$ at
this bias voltage.} \label{figm5}
\end{figure}

\begin{figure}
\centerline{\epsfig{file=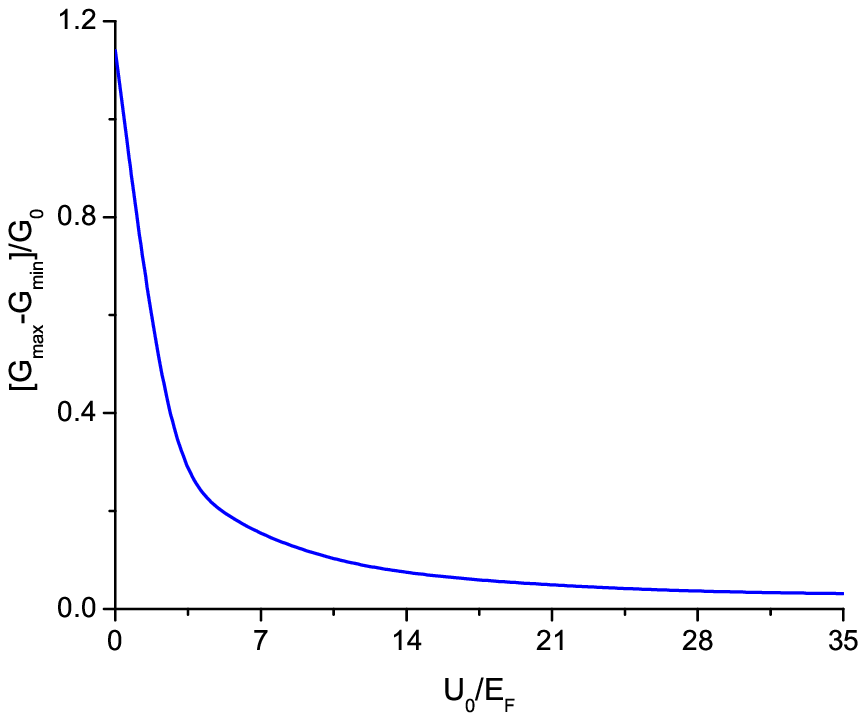,width=6cm}}
\caption{Plot of amplitude of oscillation $(G_{\rm max} -G_{\rm
min})/G_0$ of zero-bias tunneling conductance as a function of
$U_0/E_F$ for $V_0=6 E_F$ and $\Delta_0=0.01 E_F$. The oscillation
amplitudes always decay monotonically with increasing $U_0$
independent of $V_0$.} \label{figm6}
\end{figure}

Next, we consider the case $U_0 \neq 0$, so that $\Delta_0 \simeq
E_F \ll (E_F + U_0)$. In this regime, there is a large mismatch of
Fermi surfaces on the normal and superconducting sides. Such a
mismatch is well-known to act as an effective barrier for NBS
junctions. Consequently, additional barrier created by the gate
voltage becomes irrelevant, and we expect the tunneling conductance
to become independent of the applied gate voltage $V_0$. Also note
that at $eV =E_F$, $\alpha_c=0$ (Eq.\ \ref{criti1}). Hence there is
no Andreev reflection and consequently $G_0$ vanishes for all values
of the applied gate voltage for this bias voltage. Our results in
this limit, coincides with those of Ref.\ \onlinecite{beenakker1}. 
Finally in Fig.\ \ref{figm6}, we show the dependence of amplitude of
oscillation of zero-bias tunneling conductance on $U_0$ for the
applied bias voltages $V_0 =6E_F$ and $\Delta_0=0.01 E_F$. As
expected, the oscillation amplitude with decreases monotonically
with increasing $U_0$. We have verified that this feature is
independent of the applied gate voltage $V_0$ as long as $V_0 \ge
V_{0c}$.

\subsection{SBS junction} \label{SBS}
For this junction, the region \textbf{I} in Fig \ref{junction_fig} is a superconducting region
and the pair-potential can be given as \begin{eqnarray}
\Delta({\bf r}) = \Delta_0 \left[\exp(i\phi_2) \theta(x) +
\exp(i\phi_1) \theta(x+d)\right] \label{pp}
\end{eqnarray}
where $\Delta_0$ is the amplitude and $\phi_{1(2)}$ are the phases
of the induced superconducting order parameters in regions I (II) as
shown in Fig.\ref{junction_fig}, and $\theta$ is the Heaviside step
function.
Solving Eq.\ \ref{bdg1}, the wavefunctions in the
superconducting and the barriers regions are obtained. In region I,
the wavefunctions for the DBdG
quasiparticles moving along $\pm x$ direction with a transverse
momentum $k_y=q = 2\pi n/L$ (for integer $n$) and energy $\epsilon$,
are given by \cite{beenakker1}
\begin{eqnarray}
\psi_{I}^{\pm} &=& \left( u_1^{\pm}, u_2^{\pm},u_3^{\pm},u_4^{\pm}
\right) e^{\left[ i\left(\pm k_s x +q y\right) + \kappa x\right]}
 \label{supwave1}
\end{eqnarray}
where
\begin{eqnarray}
\frac{u_2^{\pm}}{u_1^{\pm}} &=& \pm \exp(\pm i\gamma), \quad
\frac{u_3^{\pm}}{u_1^{\pm}} = \exp[-i(\phi_1 \mp \beta)],\nonumber\\
\frac{u_4^{\pm}}{u_1^{\pm}} &=& \pm \exp[\pm i(\mp \phi_1 +\beta +
\gamma)], \label{ratieq1}
\end{eqnarray}
and $\sum_{i=1,4} |u_i|^2 \simeq 2\kappa$ is the normalization
condition for the wavefunction for $d \ll \kappa^{-1}$, where
$\kappa^{-1} = (\hbar v_F)^2 k_s/\left[E_F \Delta_0
\sin(\beta)\right]$ is the localization length. Here $k_s =
\sqrt{\left(E_F/\hbar v_F\right)^2 -q^2}$, $\gamma$, the angle of
incidence for the quasiparticles, is given by $\sin(\gamma) =\hbar
v_F q/E_F$, and $\beta$ is given by
\begin{eqnarray}
\beta  &=& \cos^{-1} \left(\epsilon/\Delta_0\right) \quad {\rm if}
\left|\epsilon\right| < \Delta_0 ,\nonumber\\
&=& -i \cosh^{-1} \left(\epsilon/\Delta_0\right) \quad {\rm if}
\left|\epsilon\right| > \Delta_0, \label{betaeqjc}
\end{eqnarray}
Note that for $\left|\epsilon\right| > \Delta_0$, $\kappa$ becomes
imaginary and the quasiparticles can propagate in the bulk of the
superconductor. The wavefunctions in region II ($x \ge 0$ ) can also
be obtained in a similar manner
\begin{eqnarray}
\psi_{II}^{\pm} &=& \left( v_1^{\pm}, v_2^{\pm},v_3^{\pm},v_4^{\pm}
\right) e^{\left[ i\left(\pm k_s x +q y\right) - \kappa x\right]},
 \label{supwave2}
\end{eqnarray}
where $\sum_{i=1,4} |v_i|^2=2\kappa$ and the coefficients $v_i$ are
given by
\begin{eqnarray}
\frac{v_2^{\pm}}{v_1^{\pm}} &=& \pm \exp(\pm i\gamma), \quad
\frac{v_3^{\pm}}{v_1^{\pm}} = \exp[-i(\phi_2 \pm \beta)],\nonumber\\
\frac{v_4^{\pm}}{v_1^{\pm}} &=& \pm \exp[\pm i(\mp \phi_2 -\beta +
\gamma)], \label{ratieq2}
\end{eqnarray}

The wavefunctions for electrons and holes moving along $\pm x$ in
the barrier region is given by
\begin{eqnarray}
\psi_B^{e \pm} &=& \left(1,\pm e^{\pm i \theta},0,0\right) \exp
\left[i\left(\pm k_{b} x + q y \right)\right]/\sqrt{2d}, \nonumber\\
\psi_B^{h \pm} &=&  \left(0,0,1,\pm e^{\mp i \theta'}\right) \exp
\left[i \left(\pm k'_{b} x + q y \right)\right]/\sqrt{2d}.
\label{barwave} \nonumber
\end{eqnarray}
Here the angle of incidence of the electron(hole) $\theta(\theta')$
and is given by:
\begin{eqnarray}
\sin\left[\theta(\theta')\right] &=& \frac{\hbar v_F q}{\epsilon
+(-)(E_F-V_0)} \nonumber\\
k_b (k'_b) &=& \sqrt{ \left(\frac{\epsilon +(-)(E_F-V_0)}{\hbar
v_F}\right)^2 -q^2} \label{bareq2}
\end{eqnarray}

To compute the Josephson current in the SBS junction,
the energy dispersion of the subgap Andreev bound states are found
which are localized with localization length $\kappa^{-1}$ at the barrier
\cite{zagoskin1jc,kwon1jc}. The energy dispersion $\epsilon_n$
(corresponding to the subgap state characterized by the quantum
number $n$) of these states depends on the phase difference $\phi=
\phi_2-\phi_1$ between the superconductors.
The Josephson current $I$ across the junction at a temperature $T_0$
is given by \cite{beenakker2,zagoskin1jc}
\begin{eqnarray}
I(\phi;\chi,T_0) &=& \frac{4e}{\hbar} \sum_{n} \sum_{q=-k_F}^{k_F}
\frac{\partial \epsilon_n}{\partial \phi} f(\epsilon_n),\label{jc1}
\end{eqnarray}
where $f(x)=1/(e^{x/(k_B T_0)}+1)$ is the Fermi distribution
function and $k_B$ is the Boltzman constant 
\cite{comment3}

To obtain these subgap Andreev bound states,
boundary conditions at the barrier are imposed. The wavefunctions in the
superconducting and barrier regions are constructed using Eqs.\
\ref{supwave1}, \ref{supwave2} and \ref{barwave} as
\begin{eqnarray}
\Psi_I &=& a_1 \psi_I^{+}+ b_1 \psi_{I}^{-} \quad
\Psi_{II} = a_2 \psi_{II}^{+}+ b_2 \psi_{II}^{-}, \nonumber\\
\Psi_B &=& p \psi_B^{e +}+q \psi_B^{e -} + r \psi_B^{h +} + s
\psi_N^{h -}, \label{wave2_jc}
\end{eqnarray}
where $a_1$($a_2$) and $b_1$($b_2$) are the amplitudes of right and
left moving DBdG quasiparticles in region I(II) and $p$($q$) and
$r$($s$) are the amplitudes of right(left) moving electron and holes
respectively in the barrier. These wavefunctions satisfy the
boundary conditions:
\begin{eqnarray}
\Psi_I |_{x=-d} &=& \Psi_B |_{x=-d},  \quad  \Psi_{B} |_{x=0} =
\Psi_{II} |_{x=0}. \label{bc1_jc}
\end{eqnarray}

Substituting Eqs.\ \ref{supwave1}, \ref{supwave2}, \ref{barwave},
and \ref{wave2_jc} in Eq.\ \ref{bc1_jc}, we get eight linear
homogeneous equations for the coefficients $a_{i=1,2}$, $b_{i=1,2}$,
$p$, $q$, $r$, and $s$, so that the condition for non-zero solutions
of these coefficients can be obtained as
\begin{eqnarray}
{\mathcal A'} \sin(2 \beta) + {\mathcal B'} \cos(2 \beta) +
\mathcal{C'} =0 \label{bsd1}
\end{eqnarray}
where ${\mathcal A',\,B'},\,{\rm and}\,{\mathcal C'}$ are given by
\begin{eqnarray}
{\mathcal A'} &=& \cos(k'_b d) \cos(\gamma) \cos(\theta') \sin(k_b
d)
\left(\sin(\gamma) \sin(\theta)-1\right) \nonumber\\
&& + \cos(k_b d) \cos(\gamma) \cos(\theta) \sin(k'_b d)
\nonumber\\
&& + \frac{1}{2} \cos(k_b d) \cos(\theta) \sin(2\gamma)
\sin(\theta') \sin(k'_b
d)\nonumber\\
{\mathcal B'} &=& \sin(k'_b d) \sin(k_b d) \big [-1+
\sin(\theta)\sin(\gamma) \nonumber\\
&& -\sin(\theta')\sin(\gamma) + \sin(\theta)\sin(\theta')
\sin^2(\gamma) \big] \nonumber\\
&& -\cos(k_bd) \cos(k'_b d) \cos^2(\gamma) \cos(\theta)
\cos(\theta') \nonumber\\
{\mathcal C'} &=& \cos^2(\gamma) \cos(\theta) \cos(\theta')
\cos(\phi) - \sin(k_b d) \sin(k'_b d) \nonumber\\
&& \times \left[\sin(\theta) \sin(\theta')-\sin^2(\gamma) \right.
\nonumber\\
&& \left. + \sin(\gamma) \left( \sin(\theta) -\sin(\theta') \right)
\right] \label{cf1}
\end{eqnarray}
Note that in general the coefficients ${\mathcal A'}$, ${\mathcal
B'}$, and ${\mathcal C'}$ depends on $\epsilon$ through $k_b$,
$k'_b$, $\theta$ and $\theta'$ which makes it impossible to find an
analytical solution for Eq.\ \ref{bsd1}. However, for subgap states
in graphene SBS junctions, $\epsilon \le \Delta_0 \ll E_F$. Further,
for short tunnel barrier we have $|V_0 -E_F| \ge E_F$. In this
regime, as can be seen from Eqs.\ \ref{bareq2},  ${\mathcal A'}$,
${\mathcal B'}$, and ${\mathcal C'}$ become independent of
$\epsilon$ since $k_b \simeq  k'_b \simeq k_1 = \sqrt{[(E_F -
V_0)/\hbar v_F]^2 -q^2}$ and $\theta \simeq -\theta' \simeq \theta_1
= \sin^{-1}\left[\hbar v_F q/(E_F-V_0)\right]$ so that the
$\epsilon$ dependence of $k_b$, $k'_b$, $\theta$ and $\theta'$ can
be neglected. In this regime one finds that ${\mathcal A',B',C'}
\rightarrow {\mathcal A,B,C}$ where

\begin{figure}
\centerline{\epsfig{file=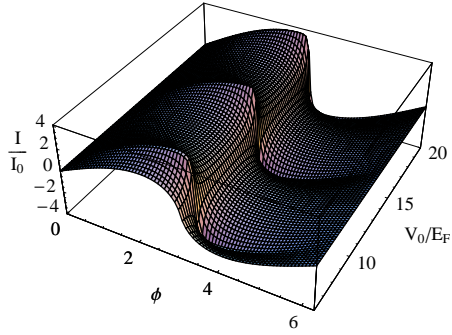, width=6cm}}
\caption{Plot of Josephson current $I$ as a function of phase
difference $\phi$ and the applied gate voltage $V_0$ for $k_B
T_0=0.01 \Delta_0$ and $d=0.5 \lambda$ showing oscillatory behavior
of $I/I_0$ as a function of the applied gate voltage.}
 \label{figm1jc}
\end{figure}
\begin{eqnarray}
{\mathcal A} &=& 0 \nonumber\\
{\mathcal B} &=& -\sin^2(k_1 d) \left[1-\sin(\gamma)\sin(\theta_1)
\right]^2 \nonumber\\
&& -\cos^2(k_1 d) \cos^2(\gamma) \cos^2(\theta_1) \nonumber\\
{\mathcal C} &=& \sin^2(k_1 d) \left[\sin(\gamma) - \sin(\theta_1)
\right]^2 \nonumber\\
&& + \cos^2(\gamma) \cos^2(\theta_1) \cos(\phi) \label{cf2}
\end{eqnarray}
The dispersion of the Andreev subgap states can now be obtained from
Eqs.\ \ref{bsd1} and \ref{betaeqjc}. There are two
Andreev subgap states with energies $\epsilon_{\pm} = \pm \epsilon$
where
\begin{eqnarray}
\epsilon &=& \Delta_0 \sqrt{1/2- {\mathcal C}/2{\mathcal B}}
\label{as1}
\end{eqnarray}
Using Eq.\ \ref{jc1}, one can now obtain the expression for the
Josephson current
\begin{eqnarray}
I(\phi,V_0,d,T_0) &=& I_0 g(\phi,V_0,d,T_0), \nonumber\\
g(\phi,V_0,d,T_0) &=& \int_{-\frac{\pi}{2}}^{\frac{\pi}{2}} d\gamma
\Bigg[ \frac{\cos^3(\gamma) \cos^2(\theta_1) \sin(\phi) }{{\mathcal
B}
\epsilon/\Delta_0} \nonumber\\
&& \times \tanh(\epsilon/2k_B T_0) \Bigg] \label{jc2a}
\end{eqnarray}
where $I_0 = e \Delta_0 E_F L/2\hbar^2 \pi v_F$ and we have replaced
$\sum_{q} \rightarrow E_F L/(2\pi \hbar v_F) \int_{-\pi/2}^{\pi/2} d
\gamma \cos(\gamma)$ as appropriate for wide junctions
\cite{beenakker2}.

\begin{figure}
\centerline{\psfig{file=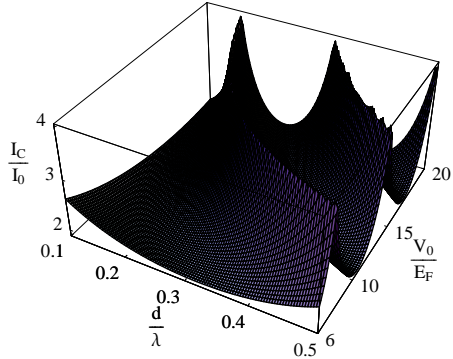,width=6cm}}
\caption{Plot of $I_c/I_0$ vs the applied gate voltage $V_0$ and the
junction thickness $d$ for $T_0=0.01 \Delta_0$.} \label{figm2jc}
\end{figure}

The dispersion of the
Andreev subgap states and the Josephson current in graphene SBS
junctions, in complete contrast to their conventional counterparts
\cite{likharev1jc,golubov1jc,zagoskin1jc}, is found to be
an oscillatory function of the
applied gate voltage $V_0$ and the barrier thickness $d$. This
statement can be most easily checked by plotting the Josephson
current $I$ as a function of the phase difference $\phi$ and the
applied gate voltage $V_0$ for a representative barrier thickness
$d=0.5 \lambda$ and temperature $k_B T_0=0.01 \Delta_0$, as done in
Fig.\ \ref{figm1jc}. In Fig.\ \ref{figm2jc}, we plot the critical
current of these junctions $I_c(V_0,d,T_0)= {\rm
Max}[I(\phi,V_0,d,T_0)]$ as a function of the applied gate voltage
$V_0$ and barrier thickness $d$ for low temperature $k_B T_0 =0.01
\Delta_0$. The critical current of these graphene SBS
junctions is an oscillatory function of both $V_0$ and $d$. This
behavior is to be contrasted with those of conventional junctions
where the critical current is a monotonically decreasing function of
both applied bias voltage $V_0$ and junction thickness $d$
\cite{likharev1jc,golubov1jc,zagoskin1jc}.

\begin{figure}
\centerline{\epsfig{file=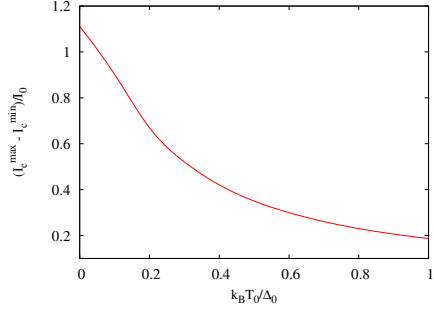, width=6cm}}
\caption{Plot of the temperature dependence of the amplitude of
oscillations of $I_c$ (given by $[I_c^{\rm max}(d)-I_c^{\rm
min}(d)]/I_0$) for $d=0.3 \lambda$. The amplitude is measured by
noting the maximum and minimum values of the critical current by
varying $V_0$ for a fixed $d$.} \label{figm3jc}
\end{figure}

\begin{figure}
\centerline{\epsfig{file=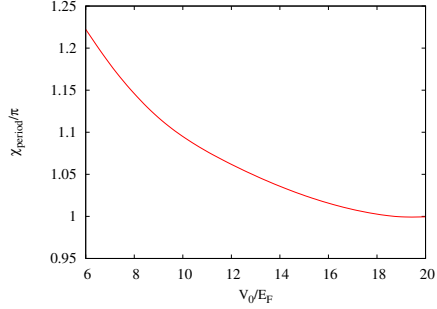,width=6cm}}
\caption{Plot of $\chi_{\rm period}$ of the critical current $I_c$
as a function of $V_0$. Note that $\chi_{\rm period}$ approaches
$\pi$ as we approach the thin barrier limit.} \label{figp1}
\end{figure}

We analyze a few other properties of these oscillations.
To find the amplitude of oscillation, we
compute $I_c$ as a function of $V_0$ (for a representative value of
$d=0.3 \lambda$), note the maximum ($I_c^{\rm max}$) and minimum
($I_c^{\rm min}$) values of $I_c$, and calculate the amplitude
$I_c^{\rm max}-I_c^{\rm min}$. The procedure is repeated for several
temperatures $T_0$ and the result is plotted in Fig.\ \ref{figm3jc}
which shows that the amplitude of oscillations decreases
monotonically as a function of temperature.
Next, we discuss the period of oscillation of the critical
current. To obtain the period, the critical current $I_c$
as a function of barrier width $d$ for the fixed applied gate
voltage $V_0$ is computed and $d_{\rm period}$ is noted down . Then
$\chi_{\rm period} = V_0 d_{\rm period}/\hbar v_F$ is computed and
$\chi_{\rm period}$ as a function of $V_0$ for $k_B T_0= 0.01
\Delta_0$ is plotted as shown in Fig.\ \ref{figp1}. It is found that $\chi_{\rm
period}$ decreases with $V_0$ and approaches an universal value
$\pi$ for large $V_0 \ge 20 E_F$. This property, as we shall see in
the next section, can be understood by analysis of graphene SBS
junctions in the thin barrier limit ($V_0 \rightarrow \infty$ and
$d\rightarrow 0$ such that $\chi= V_0 d /\hbar v_F$ remains finite
\cite{sengupta1}) and is a direct consequence of transmission
resonance phenomenon of DBdG quasiparticles in superconducting
graphene.

\subsubsection{Thin barrier limit}
\label{se2}

In the limit of thin barrier, where $V_0 \rightarrow \infty$ and
$d\rightarrow 0$ such that $\chi= V_0 d /\hbar v_F$ remains finite,
$\theta_1 \rightarrow 0$ and $k_1 d \rightarrow \chi$. From Eqs.\
\ref{cf2} and \ref{as1}, we find that in this limit, the dispersion
of the Andreev bound states becomes

\begin{eqnarray}
\epsilon_{\pm}^{\rm tb}(q,\phi;\chi)&=& \pm \Delta_0
\sqrt{1-T(\gamma,\chi)\sin^2(\phi/2)}, \label{abd}\\
T(\gamma,\chi) &=& \frac{\cos^2(\gamma)}{1-\cos^2(\chi)
\sin^2(\gamma)}. \label{teq}
\end{eqnarray}
where the superscript `tb' denote thin barrier limit. The Josephson
current $I$ can be obtained substituting Eq.\ \ref{teq} in Eq.\
\ref{jc1}. In the limit of wide junctions, one gets
\begin{eqnarray}
I^{\rm tb}(\phi,\chi,T_0) &=& I_0 g^{\rm tb} (\phi,\chi,T_0),
\nonumber\\
g^{\rm tb} (\phi,\chi,T_0)&=& \int_{-\pi/2}^{\pi/2} d \gamma \,
\Bigg[ \frac{T(\gamma,\chi) \cos(\gamma)A schemat
\sin(\phi)}{\sqrt{1-T(\gamma,\chi)\sin^2(\phi/2)}} \nonumber\\
&& \times \tanh \left(\epsilon_+ /2 k_B T_0\right) \Bigg].
\label{jc2}
\end{eqnarray}
We find that the Josephson current in
graphene SBS junctions is a $\pi$ periodic oscillatory function of
the effective barrier strength $\chi$ in the thin barrier limit.
Further we observe that the transmission probability of the DBdG
quasiparticles in a thin SBS junction is given by $T(\gamma,\chi)$
which is also the transmission probability of a Dirac quasiparticle
through a square potential barrier as noted in Ref.
\onlinecite{geim1}. Note that the transmission becomes unity for
normal incidence ($\gamma=0$) and when $\chi=n\pi$. The former
condition is a manifestation of the Klein paradox for DBdG
quasiparticles \cite{geim1}. However, this property is not reflected
in the Josephson current which receives contribution from
quasiparticles approaching the junction at all angles of incidence.
The latter condition ($\chi=n\pi$) represents transmission resonance
condition of the DBdG quasiparticles. Thus the barrier becomes
completely transparent to the approaching quasiparticles when
$\chi=n \pi$ and in this limit the Josephson current reduces to its
value for conventional tunnel junctions in the Kulik-Omelyanchuk
limit: $I^{\rm tb} (\phi,n\pi,T_0) = 4 I_0 \sin(\phi/2) {\rm
Sgn}(\cos(\phi/2)) \tanh \left(\Delta_0
\left|\cos(\phi/2)\right|/2k_B T_0\right)$ \cite{ko1jc}. This yields
the critical Josephson current $I_c^{\rm tb} (\chi=n\pi) = 4I_0$ for
$k_B T_0 \ll \Delta_0$. Note, however, that in contrast to
conventional junctions $T(\gamma,\chi)$ can not be made arbitrarily
small for all $\gamma$ by increasing $\chi$. Hence $I_c^{\rm tb}$
never reaches the Ambegaokar-Baratoff limit of conventional tunnel
junctions \cite{ambe1jc}. Instead, $I_c^{\rm tb}(\chi)$ becomes a
$\pi$ periodic oscillatory function of $\chi$. The amplitude of
these oscillations decreases monotonically with temperature.

\begin{figure}
\centerline{\epsfig{file=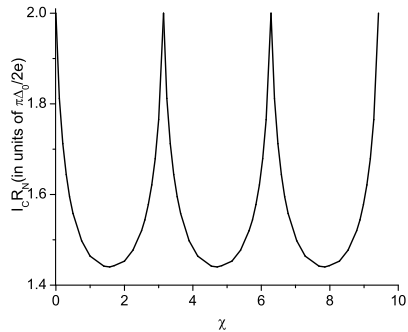, width=6cm}}
\caption{Plot of $I_c^{\rm tb} R_N$ as a function of $\chi$.
$I_c^{\rm tb} R_N$ is an oscillatory bounded function of $\chi$ and
never reaches its value ($\pi \Delta_0/2e$) for conventional
junctions in the Ambegaokar-Baratoff limit.} \label{figicrn}
\end{figure}

Finally, the product $I_c^{\rm tb} R_N$ which is
routinely used to characterize Josephson junctions
\cite{likharev1jc,golubov1jc} is computed, where $R_N$ is the normal state
resistance of the junction. For graphene SBS junctions $R_N$
corresponds to the resistance of a Dirac quasiparticle as it moves
across a normal metal-barrier-normal metal junction. For short and
wide junctions discussed here, it is given by $R_N = R_0/s_1(\chi)$
where $R_0  = \pi^2 v_F \hbar^2/(e^2 E_F L)$ and $s_1(\chi)$ is
given by \cite{geim1,beenakker2}
\begin{eqnarray}
s_1(\chi) &=& \int_{-\pi/2}^{\pi/2} d \gamma \, T(\gamma,\chi).
\cos(\gamma).
\end{eqnarray}
Note that $s_1(\chi)$ and hence $R_N$ is an oscillatory function of
$\chi$ with minimum $0.5 R_0$ at $\chi=n \pi$ and maximum $0.75 R_0$
at $\chi=(n+1/2)\pi$. The product $I_c^{\rm tb} R_N$, for thin SBS
junctions is given by
\begin{eqnarray}
I_c^{\rm tb} R_N &=& (\pi \Delta_0/2e) g^{\rm tb}_{\rm
max}(\chi,T)/s_1(\chi), \label{icrneq}
\end{eqnarray}
where $g_{\rm max}^{\rm tb} (\chi)$ is the maximum value of $g^{\rm
tb}(\phi,\chi)$. Note that $I_c^{\rm tb} R_N$ is independent of
$E_F$ and hence survives in the limit $E_F \rightarrow 0$
\cite{beenakker2}. For $k_B T_0 \ll \Delta_0$, $g_{\rm max}^{\rm tb}
(n\pi)=4$ and $s_1(n\pi)=2$, so that $I_c^{\rm tb} R_N|_{\chi=n\pi}
= \pi \Delta_0/e$ which coincides with Kulik-Omelyanchuk limit for
conventional tunnel junctions \cite{kwon1jc,ko1jc}. However, in contrast
to the conventional junction, $I_c^{\rm tb} R_N$ for graphene SBS
junctions do not monotonically decrease to the Ambegaokar-Baratoff
limit \cite{kwon1jc,ambe1jc} of $\pi \Delta_0/2e \simeq  1.57
\Delta_0/e$ as $\chi$ is increased, but demonstrates $\pi$ periodic
oscillatory behavior and remains bounded between the values $\pi
\Delta_0/e$ at $\chi=n\pi$ and $2.27 \Delta_0/e$ at
$\chi=(n+1/2)\pi$, as shown in Fig.\ \ref{figicrn}.

\section{Experiments} \label {experiments}
Superconductivity has recently been experimentally realized in
graphene \cite{heersche1}. In the proposed experiment to observe
the oscillatory behavior in the tunneling conductance and Josephson current, one needs to
realize these junctions in graphene. The local barrier can be
fabricated using methods of Refs.\ [\onlinecite{nov2},\onlinecite{zhang1}].
The easiest experimentally achievable regime corresponds to $\Delta_0 \ll E_F$
with aligned Fermi surfaces for the normal and superconducting
regions. We suggest measurement of tunneling conductance curves at
zero-bias ($eV=0$) in this regime. Our prediction is that the
zero-bias conductance will show an oscillatory behavior with the
bias voltage. In graphene, typical Fermi energy can be $E_F \le
40$meV and the Fermi-wavelength is $\lambda \geq 100$nm
\cite{geim1,nov2,zhang1,heersche1}. Effective barrier strengths of $\le 80$meV
\cite{geim1} and barrier widths of $d \simeq 10-50$ nm therefore
specifies the range of experimentally feasible junctions
\cite{geim1,nov2,zhang1}. Consequently for experimental junctions, the
ratio $V_0/E_F$ can be arbitrarily large within these parameter
ranges by fixing $V_0$ and lowering $E_F$. Experimentally, one can
set $5 \le E_F \le 20$meV so that the conditions $\Delta_0 \ll E_F$
$V_0/E_F \gg 1$ is easily satisfied for realistic $\Delta_0 \sim
0.5$meV and $V_0=200$meV. This sets the approximate range $V_0/E_F
\ge 10$ for the experiments. Note that since the period (amplitude)
of oscillations increases (decreases) as $V_0/E_F \to 1$, it is
preferable to have sufficiently large values of $V_0/E_F$ for
experimental detection of these oscillations.

To check the oscillatory behavior of the zero-bias tunneling
conductance, it would be necessary to change $V_0$ in small steps
$\delta V_0$. For barriers of a fixed width, for example with values
of $d/\lambda=0.3$, it will be enough to change $V_0$ in steps of
approximately $20-30$meV, which should be experimentally feasible.

We note that for the above-mentioned range of $V_0/E_F$, the
experimental junctions shall not always be in the thin barrier
limit. For example, as is clear from Fig.\ \ref{figm2}, the
periodicity of oscillations $\chi_{\rm period}$ of the zero-bias
tunneling conductance of such junctions shall be a function of $V_0$
and shall differ from $\pi$. This justifies the theoretical study of
NBS junctions in graphene which are away from the thin barrier
limit.

Apart from the above-mentioned experiments, it should also be
possible to measure the tunneling conductance as a function of the
applied bias voltage $eV/ \Delta_0$ for different applied gate
voltages $V_0$. Such measurements can be directly compared with
Fig.\ \ref{figm3}. Finally, it might be also possible to create a
relative bias $U_0$ between the Fermi surfaces in the normal and
superconducting side and compare the dependence of oscillation
amplitudes of zero-bias tunneling conductance on $U_0$ with the
theoretical result shown in Fig.\ \ref{figm5}.

We also suggest measuring DC Josephson
current in these junctions as a function of the applied voltage
$V_0$. Such experiments for conventional Josephson junctions are
well-known \cite{ar1jc}. Further SNS junctions in graphene has also
been recently been experimentally created \cite{heersche1,ref9}.
To observe the oscillatory behavior of the Josephson current, alike the
procedure to measure the tunneling conductance,
it would be necessary to change $V_0$ in small steps $\delta V_0$.
For barriers with fixed $d/\lambda=0.3$ and $V_0/E_F=10$, this would
require changing $V_0$ in steps of approximately $30$meV which is
experimentally feasible. The Joule heating in such junctions,
proportional to $I_c^2 R_N$, should also show measurable oscillatory
behavior as a function of $V_0$.
\section{Kondo effect and STM Spectra}

\subsection{Kondo effect in Graphene}
\label{Kondosec}

In this section, we shall present a large $N$ analysis for a generic
local moment coupled to Dirac electrons in graphene to show that
Kondo effect in graphene is unconventional can be tuned by gate
voltage. We demonstrate the presence of a finite critical Kondo
coupling strength in neutral graphene. We point out that local
moments in graphene can lead to non Fermi-liquid ground state via
multi channel Kondo effect.

The crucial requirement for occurrence of Kondo effect is that the
embedded impurities should retain their magnetic moment in the
presence of conduction of electrons of graphene. We will not
quantitatively address the problem of local moment formation in the
presence of Dirac sea of electrons in graphene in the present paper.
We expect that large band width and small linearly vanishing density
of states at the fermi level in graphene should make survival of
impurity magnetic moment easier than in the conventional 3D metallic
matrix. A qualitative estimate of the resultant Kondo coupling can
be easily made considering hybridization of electrons in $\pi$ band
in graphene with $d$ orbitals of transition metals. Typical hopping
matrix elements for electrons in $\pi$ band is $t \sim 2$eV and
effective Hubbard $U$ in transition metals is $8$eV. So the Kondo
exchange $J \sim 4t^2/U$, estimated via standard Schrieffer-Wolf
transformation, can be as large as $2$ eV which is close to one of
the largest $J \simeq 2.5$ eV for Mn in Zn. Therefore it is customary
to use Kondo Hamiltonian \cite{kondoref1} to study the effect in Graphene.

\subsubsection{Large N analysis}\label{large_N}
Our analysis begins with the Hamiltonian for non-interacting Dirac
electron in graphene. In the presence of a gate voltage $V$, the
Hamiltonian can be expressed in terms of electron annihilation
operators $\Psi_{A(B) \alpha}^s$ at sublattice $A(B)$ and Dirac
point $s=K,K$ with spin $\alpha=\uparrow,\downarrow$ as
\begin{eqnarray}
H &=& \int \frac{d^2 k}{(2\pi)^2} \left( \Psi_{A \alpha}^{s \dagger}({\bf k}),
\Psi_{B \alpha}^{s\dagger}({\bf k})\right) \nonumber\\
&& \times \left(\begin{array}{cc} {eV \quad \hbar v_F (k_x-i {\rm sgn}(s) k_y)} \\
{\hbar v_F(k_x+i {\rm sgn}(s) k_y) \quad  eV} \end{array}\right)
\left(
\begin{array}{c} {\Psi_{A \alpha}^{s}({\bf
k})}\\{\Psi_{B\alpha}^{s}({\bf k})}
\end{array}\right) \label{ke1}
\end{eqnarray}
where ${\rm sgn}(s)=1(-1)$ for $s=K(K')$, $v_F$ is the Fermi
velocity of graphene, and all repeated indices are summed over. In
Eq.\ \ref{ke1} and rest of the analysis, we shall use an upper
momentum cutoff $k_c = \Lambda/(\hbar v_F)$, where $\Lambda \simeq
2$eV corresponds to energy up to which the linear Dirac dispersion
is valid, for all momenta integrals.

Eq.\ \ref{ke1} can be easily diagonalized to obtain the eigenvalues
and eigenfunctions of the Dirac electrons: $E_{\pm} = eV \pm \hbar
v_F k$ where ${\bf k} = (k_x,k_y)=(k,\theta)$ denote momenta in
graphene and $(u_A^{s \pm},u_B^{s \pm}) = 1/{\sqrt{2}} \left(1, \pm
\exp \left(i {\rm sgn}(s) \theta\right) \right)$. Following Ref.\
\onlinecite{frad1}, we now introduce the $\xi$ fields, which represents
low energy excitations with energies $E_{\pm}$, and write
\begin{eqnarray}
\Psi_{A\alpha}^{s}({\bf k}) &=& \sum_{j=\pm} u^{sj}_A
\xi_{j\alpha}^s = 1/\sqrt{2} (\xi_{+ \alpha}^s ({\bf k})+ \xi_{-
\alpha}^s ({\bf k})), \nonumber\\
\Psi_{B\alpha}^{s}({\bf k}) &=& \exp(i\theta)/\sqrt{2} (\xi_{+
\alpha}^s ({\bf k}) - \xi_{- \alpha}^s ({\bf k})).   \label{xifef}
\end{eqnarray}

In what follows, we shall consider a single impurity to be centered
around ${\bf x}=0$. Thus to obtain an expression for the coupling
term between the local moment and the conduction electrons, we shall
need to obtain an expression for $\Psi({\bf x}=0)\equiv \Psi(0)$. To
this end, we expand the $\xi$ fields in angular momentum channels
$\xi_{+ \alpha}^s ({\bf k})=\sum_{m=-\infty}^{\infty} e^{i m \theta}
\xi_{+ \alpha}^{m s} (k)$, where we have written ${\bf
k}=(k,\theta)$. After some straightforward algebra, one obtains
\begin{eqnarray}
\Psi_{B\alpha}^{s}(0)&=& \frac{1}{\sqrt{2}} \int_0^{k_c} \frac{k
dk}{2\pi} \left( \xi_{+ \alpha}^{-{\rm sgn}(s) s} (k) -
\xi_{- \alpha}^{-{\rm sgn}(s) s} (k)\right), \nonumber\\
\Psi_{A\alpha}^{s}(0)&=& \frac{1}{\sqrt{2}} \int_0^{k_c} \frac{k
dk}{2\pi} \left( \xi_{+ \alpha}^{0 s} (k) + \xi_{- \alpha}^{0 s}
(k)\right). \label{psif}
\end{eqnarray}
Note that $\Psi_B(0) $ receives contribution from $m=\pm 1$ channel
while for $\Psi_A(0)$, the $m=0$ channel contributes. The Kondo
coupling of the electrons with the impurity spin is given by
\begin{eqnarray} H_K &=& \frac{g}{2k_c^2} \sum_{s=1}^{N_s}
\sum_{l=1}^{N_f} \sum_{\alpha,\beta =1}^{N_c}  \sum_{a=1}^{N_c^2-1}
\Psi^{s \,\dagger}_{l \alpha}(0) \tau_{\alpha \beta}^a  \Psi^{s}_{l
\beta}(0) S^a, \label{coupling1}
\end{eqnarray}
where $g$ is the effective Kondo coupling for energy scales up to
the cutoff $\Lambda$, ${\bf S}$ denotes the spin at the impurity
site, ${\bf \tau}$ are the generators of the SU($N_c$) spin group,
and we have now generalized the fermions, in the spirit of large $N$
analysis, to have $N_s$ flavors (valley indices) $N_f$ colors
(sublattice indices) and $N_c$ spin. For realistic systems
$N_f=N_c=N_s=2$. Here we have chosen Kondo coupling $g$ to be
independent of sublattice and valley indices. This is not a
necessary assumption. However, we shall avoid extension of our
analysis to flavor and/or color dependent coupling term for
simplicity. Also, the Dirac nature of the graphene conduction
electrons necessitates the Kondo Hamiltonian to mix $m=\pm 1$ and
$m=0$ channels (Eqs.\ \ref{psif} and \ref{coupling1}). This is in
complete contrast to the conventional Kondo systems where the Kondo
coupling involves only $m=0$ angular momentum channel.

The kinetic energy of the Dirac electrons can also be expressed in
terms of the $\xi$ fields:
\begin{equation}
H_0 = \int_0^{\infty} \frac{k dk}{2\pi}
\sum_{m=-\infty}^{\infty} \sum_{s,\alpha} \left(E_{+}(k)
\xi_{+\alpha}^{m s \, \dagger} \xi_{+ \alpha}^{m s} + E_{-}(k)
\xi_{-\alpha}^{m s\, \dagger} \xi_{- \alpha}^{m s} \right)
\end{equation}
Typically such a term involves all angular momenta channels. For our
purpose here, it will be enough to consider the contribution from
electrons in the $m=0,\pm 1$ channels which contribute to scattering
from the impurity (Eqs.\ \ref{psif} and \ref{coupling1}). To make
further analytical progress, we now unfold the range of momenta $k$
from $(0, \infty)$ to $(-\infty,\infty)$ by defining the fields
$c_{1(2) \alpha}^{s}$
\begin{eqnarray}
c_{1(2) \alpha}^{s}(k) &=& \sqrt{\left|k\right|}
\xi_{+ \alpha}^{0(-{\rm sgn}(s)) s} (|k|), \quad k>0 ,\nonumber\\
c_{1(2) \alpha}^{s}(k) &=& +(-) \sqrt{\left|k\right|} \xi_{-
\alpha}^{0(-{\rm sgn}(s)) s} (|k|), \quad k < 0, \label{cfield}
\end{eqnarray}
so that one can express the $\Psi$ fields as $\Psi_{A(B)
\alpha}^s(0) = \int_{-\infty}^{\infty} \frac{ dk}{2\pi}
\sqrt{\left|k\right|} c_{1(2) \alpha}^s (k)$. In terms of the
$c_{1(2) \alpha}^{s}$ fields, the kinetic energy (in the $m=0,\pm 1$
channels) and the Kondo terms in the Hamiltonian can therefore be
written as
\begin{eqnarray}
H_0 &=&  \int_{-k_c}^{k_c} dk/(2\pi)  E_k
c_{l \alpha}^{s\, \dagger} c_{l \alpha}^{s} \nonumber\\
H_K &=& g/(8 \pi^2 k_c^2) \int_{-k_c}^{k_c} \int_{-k_c}^{k_c}
\sqrt{\left|k\right|} \sqrt{\left|k'\right|}
dk dk' \nonumber\\
&& \times \left( c_{l \alpha}^{s\, \dagger} (k) \, \tau_{\alpha
\beta}^a \, c_{l \beta }^{s}(k') \, S^a \right), \label{coupling2}
\end{eqnarray}
where $E_k = eV + \hbar v_F k$ and summation over all repeated
indices are assumed.
\begin{figure}
\centerline{\psfig{file=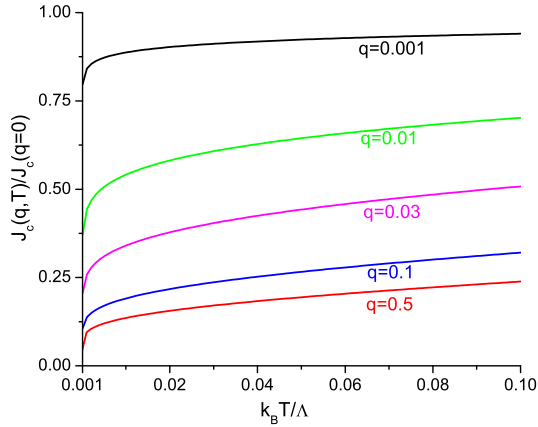,width=8cm}}
\caption{Sketch of the critical Kondo coupling $J_c(q,T)$ as a
function of temperature for several applied voltages $q=eV/\Lambda$.
The Kondo phase exists for $J > J_c$.} \label{fig1}
\end{figure}

We follow standard procedure \cite{newns1} of representing the
local spin by SU($N_c$)Fermionic fields $f_{\alpha}$ and write the
partition function of the system in terms of the $f$ and $c$ fields
\begin{eqnarray}
Z &=& \int {\mathcal D} c {\mathcal D} c^{\dagger}  {\mathcal D} f
{\mathcal D} f^{\dagger}
{\mathcal D} \epsilon  \, e^{-S/\hbar}, \quad S = S_0 + S_1 + S_2 \nonumber \\
S_0 &=& \int_0^{\beta\hbar } d \tau \int_{-k_c}^{k_c} dk/(2 \pi)
\left( c_{l \alpha}^{s\, \dagger}(k,\tau)
G_0^{-1}  c_{l \alpha}^{s} (k,\tau)\right), \nonumber\\
S_1 &=& J/(4 \pi^2 N_c k_c^2) \int_0^{\beta \hbar} d \tau
\int_{-k_c}^{k_c} \int_{-k_c}^{k_c} \sqrt{\left|k\right|}
\sqrt{\left|k'\right|} dk dk' \nonumber\\
&& \times \left[c_{l \alpha}^{s\, \dagger} (k,\tau) \, \tau_{\alpha
\beta}^a \, c_{l \beta }^{s}(k',\tau) f_{\gamma}^{\dagger}(\tau)
\tau_{\gamma \delta}^a f_{\delta}(\tau) \right] \nonumber \\
S_2 &=& \int_0^{\beta \hbar} d\tau  \left[\left(
f_{\alpha}^{\dagger}(\tau)\left[\hbar \partial_{\tau}+\epsilon(\tau)
\right] f_{\alpha} (\tau) \right)- \epsilon(\tau) Q \right],
\label{s2}
\end{eqnarray}
where $G_0^{-1} = \hbar \partial_{\tau} +E_k$ is the propagator for
$c$ fields, $J= g N_c/2$ is the renormalized Kondo coupling, we have
imposed the impurity site occupancy constraint $\sum_{\alpha}
f_{\alpha}^{\dagger} f_{\alpha} = Q $ using a Lagrange multiplier
field $\epsilon(\tau)$.

We now use the identity\cite{newns1}
\begin{equation}
\tau_{\alpha \beta}^a \tau_{\gamma
\delta}^a = N_c \delta_{\alpha \delta} \delta_{\beta \gamma} -
\delta_{\alpha \beta} \delta_{\gamma \delta}
\end{equation}
and
decouple $S_1$ using a Hubbard-Stratonovitch field $\phi_l^s$. In
the large $N_c$ limit one has $S= S_0 + S_2 + S_3 + S_4$, where
\begin{eqnarray}
S_3 &=& \int_0^{\beta \hbar} d \tau \int_{-k_c}^{k_c}
\frac{\sqrt{\left|k\right|} dk }{(2\pi)} \left(
\phi_l^{\ast\,s}(\tau) c_{l \alpha}^{s\, \dagger} (k,\tau)
f_{\alpha}(\tau) + {\rm h.c} \right)
\nonumber \\
S_4 &=& N_c k_c^2/J  \int_0^{\beta \hbar} d \tau
\phi_l^{\ast\,s}(\tau)  \phi_l^s(\tau) . \label{s4}
\end{eqnarray}
Note that at the saddle point level $ \left<\phi_l^s\right>  \sim
\left< \sum_{\alpha} c_{l\alpha} ^{s\,\dagger} f_{\alpha}\right>$ so
that a non-zero value of $\phi_l^s$ indicates the Kondo phase. In
what follows, we are going to look for the static saddle point
solution with $\phi_l^s (\tau) \equiv \phi_0$ and $\epsilon(\tau)
\equiv \epsilon_0$ \cite{newns1}. In this case, it is easy to
integrate out the $c$ and $f$ fields, and obtain an effective action
in terms of $\phi_ 0$ and $\epsilon_0$ and one gets $S' = S_5 + S_6$
with
\begin{eqnarray}
S_5 &=& -\beta \hbar N_c {\rm Tr} \left[\ln \left(i \hbar \omega_n
-\epsilon_0 - N_s N_f \phi_0^{\ast} G'_0(i\omega,V) \phi_0 \right)
\right],
\nonumber \\
S_6 &=& \beta \hbar \left( N_s N_c N_f k_c^2 \left|\phi_0\right|^2/J
- \epsilon_0 Q \right), \label{s6}
\end{eqnarray}
where ${\rm Tr}$ denotes Matsubara frequency sum as well as trace
over all matrices and the Fermion Green function $G'_0(ip_n,q)
\equiv G'_0$ is given by \cite{frad1}
\begin{eqnarray}
G'_0 &=& \frac{-\Lambda}{2 \pi (\hbar v_F)^2} (ip_n-q) \ln
\left[1/\left|ip_n-q\right|^2\right], \label{g0}
\end{eqnarray}
where, in the last line we have switched to dimensionless variables
$p_n=\hbar \omega_n/\Lambda$ and $q = eV/\Lambda$.

One can now obtain the saddle point equations from Eq.\ \ref{s6}
which are given by $\delta S'/\delta \phi_0 =0$ and $\delta
S'/\delta \epsilon_0 =0$. Using Eqs.\ \ref{s6} and \ref{g0}, one
gets (after continuing to real frequencies and for $T=0$)
\begin{eqnarray}
1/J &=& - \Lambda/(\pi \hbar v_F k_c^2 )^2 \int_{-1}^{0} dp\,
G_0(p - \nu - \Delta_0 G_0 /2)^{-1}, \nonumber\\
Q/N_c &=& 1/(2 \pi) \int_{-1}^{0} dp\, \nu (p - \nu - \Delta_0 G_0
/2)^{-1}, \label{saddle}
\end{eqnarray}
where we have defined the dimensionless variable
$\Delta_0 = N_f N_s
|\phi_0|^2/(\pi \hbar^2 v_F^2)$, $p = \hbar \omega/\Lambda$, $G_0 =
2 \pi (\hbar v_F)^2 G'_0/\Lambda$, $\nu= \epsilon_0/\Lambda \ge 0$,
and have used the energy cutoff $\Lambda$ for all frequency
integrals. At the critical value of the coupling strength, putting
$\nu=0$ and $\Delta_0=0$, we finally obtain the expression for
$J_c(q,T)$
\begin{eqnarray}
J_c(q,T) &=& J_c(0) \left[1-2q\ln\left(1/q^2\right) \ln\left(k_B
T/\Lambda\right) \right]^{-1} \label{criticalJ}
\end{eqnarray}
where the temperature $k_B T$ is the infrared cutoff, $J_c(0) = (\pi
\hbar v_F k_c^2)^2/\Lambda = \pi^2 \Lambda$ is the critical coupling
in the absence of the gate voltage, and we have omitted all
subleading non-divergent term which are not important for our
purpose. For $V=0=q$, we thus have, analogous to the Kondo effect in
flux phase systems \cite{frad1}, a finite critical Kondo coupling
$J_c(0) = \pi^2 \Lambda \simeq 20$eV which is a consequence of
vanishing density of states at the Fermi energy for Dirac electrons
in graphene. Of course, the mean-field theory overestimates $J_c$. A
quantitatively accurate estimate of $J_c$ requires a more
sophisticated analysis which we have not attempted here.

\subsubsection{Results and Discussions}
The presence of a gate voltage leads to a Fermi surface and
consequently $J_c(q,T) \rightarrow 0$ as $T \rightarrow 0$. For a
given experimental coupling $J < J_c(0)$ and temperature $T$, one
can tune the gate voltage to enter a Kondo phase. Fig.\ \ref{fig1},
which shows a plot of $J_c(q,T)$ as a function of $T$ for several
gate voltages $q$ illustrates this point. The temperature
$T^{\ast}(q)$ below which the system enters the Kondo phase for a
physical coupling $J$ can be obtained using $J_c(q,T^{\ast})=J$
which yields
\begin{eqnarray}
k_B T^{\ast} &=& \Lambda
\exp\left[(1-J_c(0)/J)/(2q\ln[1/q^2])\right]
\end{eqnarray}
For a typical $J \simeq 2$eV and voltage $eV \simeq 0.5$eV,
$T^{\ast} \simeq 35$K 
\cite{comment4}
We stress that even with
overestimated $J_c$, physically reasonable $J$ leads to
experimentally achievable $T^{\ast}$ for a wide range of
experimentally tunable gate voltages.

We now discuss the possible ground state in the Kondo phase
qualitatively. In the absence of the gate voltage a finite $J_c$
implies that the ground state will be non-Fermi liquid as also noted
in Ref.\ \onlinecite{frad1} for flux phase systems. In view of the large
$J_c$ estimated above, it might be hard to realize such a state in
undoped graphene. However, in the presence of the gate voltage, if
the impurity atom generates a spin half moment and the Kondo
coupling is independent of the valley(flavor) index, we shall have a
realization of two-channel Kondo effect in graphene owing to the
valley degeneracy of the Dirac electrons. This would again lead to
overscreening and thus a non Fermi-liquid like ground state
\cite{affleck1}. The study of details of such a ground state
necessitates an analysis beyond our large $N$ mean-field theory. To
our knowledge, such an analysis has not been undertaken for Kondo
systems with angular momentum mixing. In this work, we shall be
content with pointing out the possibility of such a multichannel
Kondo effect in graphene and leave a more detailed analysis as an
open problem for future work.

Next, we discuss experimental observability of the Kondo phenomena
in graphene. The main problem in this respect is creation of local
moment in graphene. There are several routes to solving this
problem. i) Substitution of a carbon atom by a transition metal
atom. This might in principle frustrate the strong sp$^2$ bonding
and thus locally disturb the integrity of graphene atomic net.
However, nature has found imaginative ways of incorporating
transition metal atoms in p-$\pi$ bonded planar molecular systems
such as porphyrin \cite{por1}. Similar transition metal atom
incorporation in extended graphene, with the help of suitable
bridging atoms, might be possible. ii) One can try chemisorption of
transition metal atoms such as Fe on graphene surface through sp-d
hybridization in a similar way as in intercalated graphite
\cite{chem1}. iii) It might be possible to chemically bond molecules
or free radicals with magnetic moment on graphene surface as
recently done with cobalt pthalocyanene (CoPc) molecule on AU(111)
surface \cite{copc}. This might result in a strong coupling between
graphene and impurity atom leading to high Kondo temperatures as
seen for CoPc on AU(111) surface ($T_K \simeq 280K$). iv) Recently
ferromagnetic cobalt atom clusters with sub nano-meter size,
deposited on carbon nanotube, have exhibited Kondo
resonance\cite{cluster1}. Similar clusters deposition in graphene
might be a good candidate for realization of Kondo systems in
graphene. v) From quantum chemistry arguments, a carbon vacancy, or
substitution of a carbon atom by a boron or nitrogen might lead to a
spin-half local moment formation. In particular, it has been shown
that generation of local defects by proton irradiation can create
local moments in graphite \cite{defect1}. Similar irradiation
technique may also work for graphene.

For spin one local moments and in the presence of sufficiently large
voltage and low temperature, one can have a conventional Kondo
effect in graphene. The Kondo temperature for this can be easily
estimated using $k_BT_K \sim D \exp(-1/\rho J )$ where the band
cutoff $D \simeq10$eV, $J \simeq 2-3$eV  and DOS per site in
graphene $\rho \simeq 1/20$ per eV. This yield $T_K \simeq 6-150$K.
The estimated value of $T_K$ has rather large variation due to
exponential dependence on $J$. However, we note that Kondo effect
due to Cobalt nano-particle in graphitic systems such as carbon
nanotube leads to a high $T_K \approx 50 K$ which means that a large
$J$ may not be uncommon in these systems.
Recently, It has also been shown that the Kondo effect
can be controlled by orbital degrees of freedom\cite{wehling}.
A symmetry class of orbitals in a magnetic adatoms with
inner shell in graphene leads to a distinct quantum critical points,
where $T_K\propto(J-J_c)^{1/3}$ near the critical coupling
$J_c$\cite{uchoa3}.

Finally, we note that recent experiments have shown a striking
conductance changes in carbon nanotubes and graphene, to the extent
of being able to detect single paramagnetic spin-half NO$_2$
molecule \cite{novo1}. This has been ascribed to conductance
increase arising from hole doping (one electron transfer from
graphene to NO$_2$). Although Kondo effect can also lead to
conductance changes, in view of the fact that a similar effect has
been also seen for diamagnetic NH$_3$ molecules, the physics in
these experiments is likely to be that of charge transfer and not
local moment formation.

\subsection{STM spectra of Graphene}\label{stmsection}

\begin{figure}
\centerline{\psfig{file=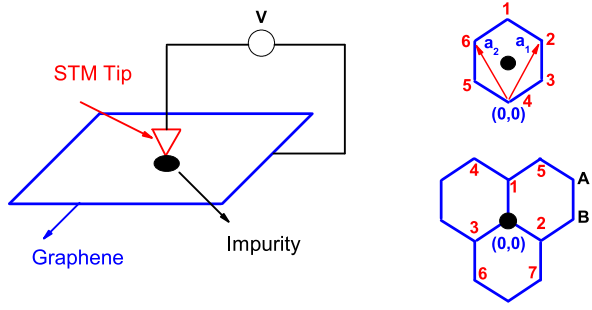,width=8cm}}
\caption{Schematic experimental setup with the right
panel showing two possible positions (atop hexagon center and atop a
B site) of the impurity. The numbers denote nearest neighbor $A$ and
$B$ sublattice sites to the impurity. $a_{1(2)}= +(-) \sqrt{3}/2
\hat x + 3/2 \hat y$ [lattice spacing set to unity] are graphene
lattice vectors. The choice of coordinate center (0,0) are shown for
each case}
\label{ra_fig1}
\end{figure}

The experimental situation for STM measurement is  schematically
represented in Fig.\ \ref{ra_fig1}. The STM tip is placed atop the
impurity and the tunneling current ${\mathcal I}$ is measured as a
function of applied bias voltage $V$. The possible positions of the
impurity is shown in the right panel of Fig.\ \ref{ra_fig1}. Such a
situation can be modeled by the well-known Anderson Hamiltonian
\cite{anderson1}. Here we incorporate the low-energy Dirac
quasiparticles of graphene in this Hamiltonian which is given by

\begin{eqnarray}
H &=& H_G + H_d+ H_{t} + H_{Gd} + H_{Gt} + H_{dt} \label{hamil1}
\\
H_G &=& \int_k \psi_s^{\beta \dagger}(\vec k)\left [\hbar v_F
(\tau_z \sigma_x k_x + \sigma_y k_y) -E_F I \right]
\psi_s^{\beta}(\vec
k) \label{hamilg} \nonumber\\
H_d &=& \sum_{s=\uparrow,\downarrow} \epsilon_d d_{s}^{\dagger}
d_{s} + U n_{\uparrow}n_{\downarrow}
\label{hamilimp} \nonumber\\
H_{t} &=& \sum_{\nu} \Big[ \sum_{\sigma=\uparrow \downarrow}
\epsilon_{t\nu} t_{\nu s}^{\dagger} t_{\nu s} +(\Delta_0 t_{\nu
\uparrow}^{\dagger} t_{-\nu \downarrow}^{\dagger}+ {\rm h.c}
)\Big] \label{hamiltip} \\
H_{Gd} &=& \sum_{\alpha=A,B} \int_k   \left(V^0_{\alpha} (\vec k)
c_{\alpha,s}^{\beta}(\vec k) d_{s}^{\dagger} + {\rm h.c.} \right)
\label{graimp}\\
H_{dt}&=& \sum_{s=\uparrow,\downarrow;\nu} \left( W^0_{\nu} t_{\nu
s} d_{s}^{\dagger}+{\rm h.c.} \right). \label{imptip}\\
H_{Gt} &=& \sum_{\alpha=A,B;\nu} \int_k \left(U^0_{\alpha;\nu} (\vec
k) c_{\alpha,s}^{\beta}(\vec k) t_{\nu s}^{\dagger} + {\rm h.c.}
\right) \label{gratip}
\end{eqnarray}

Here $H_G$ is the Dirac Hamiltonian for the graphene which
are described by the two component annihilation operator
$\psi^{\beta}_s (\vec k)= (c_{A s}^{\beta}(\vec k), c_{B
s}^{\beta}(\vec k))$ belonging to the valley $\beta=K,K'$ and spin
$s=\uparrow,\downarrow$, $I$ is the identity matrix, $\tau$ and
$\sigma$ denotes Pauli matrices in valley and pseudospin spaces,
$v_F$ is the Fermi velocity, and $\int_k \equiv \sum_{\beta= K,K'}
\sum_{s=\uparrow \downarrow} \int \frac{d^2k}{(2\pi)^2}$.
 $H_d$ denotes the impurity atom Hamiltonian with an on-site energy
$\epsilon_d$ and $U$ is the strength of on-site Hubbard interaction.
$H_t$ is the Hamiltonian for the superconducting ($\Delta_0\ne 0$)
or metallic ($\Delta_0=0$) tip electrons with on-site energy
$\epsilon_{t\nu}$ where $\nu$ signifies all quantum numbers (except
spin) associated with the tip electrons. The operators $d_{s}$ and
$t_{\nu s}$ are the annihilation operators for the impurity and the
tip electrons. The Hamiltonians $H_{Gd}$, $H_{Gt}$, and $H_{dt}$
describe interaction between the graphene and the impurity
electrons, the graphene and the STM tip electrons, and the impurity
and the STM tip electrons respectively. The corresponding
interactions parameters $V^0_{\alpha}(\vec k)$,
$U^0_{\alpha;\nu}(\vec k)$, and $W^0_{\nu}$ are taken to be
independent of valley and spin indices of graphene electrons but may
depend on their sublattice index or pseudospin. In defining the
Hamiltonian Eq.\ \ref{hamil1}, we omit inter-valley scattering of
electrons by the impurity. This is usually justified if the impurity
radius is larger than $|\vec K-\vec K'|^{-1}$ so that the
inter-valley scattering is suppressed. In the present case, for
impurity atom atop the hexagon center, the impurity scattering
potential respects the pseudospin symmetry (since it does not
distinguish between $A$ or $B$ sites) and hence can not flip
pseudospin of graphene electrons. Thus a graphene electron in
momentum state ${\vec k}_1$ around $K$ valley can only be scattered
to around momentum states ${\vec k}_2=-{\vec k}_1$ in $K'$ valley.
This constraint further reduces the phase space for inter-valley
scattering. This phase space constraint makes the inter-valley
scattering small for all impurities atop hexagon center. For
impurities atop graphene sites, our analysis is applicable for large
impurity size where inter-valley scattering is negligible.

\subsubsection{Tunneling Current}\label{tcurrent}
The tunneling current for the present model is given by
\begin{eqnarray}
 {\mathcal
I}(t)= e\langle dN_t/dt \rangle =ie \langle [H,N_t]\rangle/\hbar
\end{eqnarray}
where $N=\sum_{\nu s} t_{\nu s}^{\dagger} t_{\nu s}$ is the number
operator for the tip electrons. These commutators receive
contribution from $H_{dt}$ and $H_{Gt}$ in Eqs.\ \ref{imptip} and
\ref{gratip} and can be evaluated by a straightforward
generalization of method outlined in Ref.\ \onlinecite{wingreen1} to
the case of superconducting tips. A standard calculation yields
\begin{eqnarray}
{\mathcal I} = {\mathcal I}_0
\int_{-\infty}^{\infty}{d\omega}[f(\omega-eV)-f(\omega)]
\rho_t(\omega-eV) \Big[ \rho_G(\omega) \nonumber\\
\times |U^{0}|^2 + \frac{|B(\omega)|^2}{{\rm
Im}\Sigma_d(\omega)}\frac {|q(\omega)|^2-1+2{\rm Re}[q(\omega)]
\chi(\omega)}{(1+\chi^2(\omega))(1+\xi^2)}\Big ] \label{cur1}
\end{eqnarray}
where ${\mathcal I}_0 = 2e(1+\xi^2)/h$, $\rho_G(\epsilon)$ and
$\rho_t(\epsilon)$ are the graphene and STM tip electron DOS
respectively, $\xi = |U^0_B|/|U^0_A|= |V^0_B|/|V^0_A|$ is the ratio
of coupling of the impurity to the electrons in $B$ and $A$ sites of
graphene with $U_A^0=U^0$ and $V_A^0=V^0$, $\chi(\epsilon) =
[\epsilon -\epsilon_d - {\rm Re} \Sigma_d(\epsilon)]/{\rm Im}
\Sigma_d(\epsilon)$,$f(\epsilon) =1/(1 + \exp[\epsilon/T])$
($k_B=1$) is the Fermi function, and $\Sigma_d(\epsilon)$ is the
impurity advanced self-energy in the absence of the tip. Here
$B(\epsilon)= V^0 U^0 I_2(\epsilon)$ and $q(\epsilon)$ is given by
\begin{align}
q(\epsilon)=&\frac{W^0/U^0 + V^0 I_1(\epsilon)}{V^0 I_2(\epsilon)}
\label{qeq}
\end{align}
where,
\begin{align}
I_1(\epsilon)=& (1+\xi^2) \sum_{k} {\rm Tr} \left[ {\rm
Re}\{{\mathcal G}(\epsilon,{\bf k})\}\right]\\
I_2(\epsilon) =&
(1+\xi^2)\sum_k {\rm Tr} \left[{\rm Im}\{{\mathcal G}(\epsilon,{\bf
k})\}\right]
\end{align}
Here, we have neglected the energy dependence of the coupling
functions assuming small applied voltages.  
${\rm Tr}$ denotes trace over Pauli matrices in
pseudospin, valley and spin spaces, and ${\mathcal G}$ is the Green
function for the graphene electrons: 
\begin{equation}
{\mathcal G}(\epsilon,{\bf k})
= \frac{(\epsilon +E_F)I -\hbar v_F (\tau_z \sigma_x k_x + \sigma_y
k_y)}{(\epsilon+ E_F)^2 -\hbar^2v_F^2 |{\bf k}|^2 - i\eta}
\end{equation}
A simple calculation yields \cite{wingreen1, neto2}
\begin{eqnarray}
I_1(\epsilon) &=& -4(1+\xi^2)(\epsilon+E_F)\ln \left|1-
\Lambda^2/(\epsilon + E_F)^2 \right|/\Lambda^2 \nonumber\\
I_2(\epsilon) &=& 4(1+\xi^2) \pi |\epsilon+ E_F|
\theta(\Lambda-\epsilon - E_F ) /\Lambda^2. \label{i1i2}
\end{eqnarray}
where $\Lambda$ is the ultraviolet momentum cutoff and $\theta$ is
the Heaviside step function. Usually, in graphene, $\Lambda$ is
taken to be the energy at which the graphene bands start bending
rendering the low-energy Dirac theory inapplicable and can be
estimated to be $1-2$eV \cite{baskaran1}.

\subsubsection{Results and Discussions}
\begin{figure}
\centerline{\epsfig{file=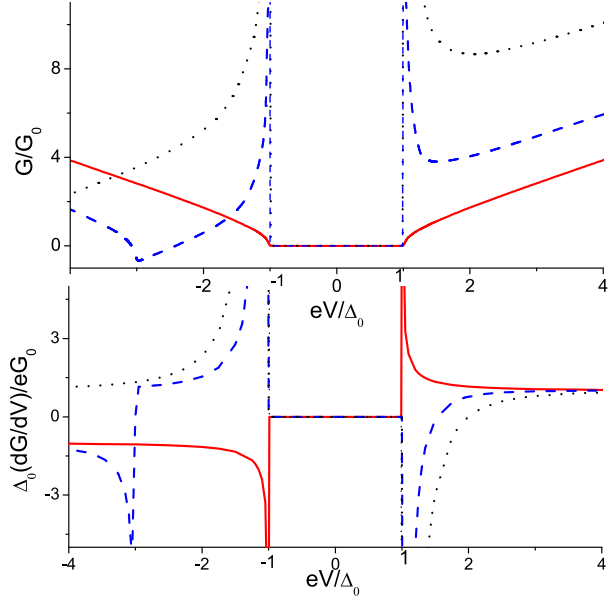,width=8cm}}
\caption{(Color online) Plot of the tunneling conductance $G$ and
its derivative $dG/dV$ as a function of the applied bias voltage
$eV/\Delta_0=-p$ for $r=0,2,6$ (red solid, blue dashed and black
dotted lines) respectively. See text for details.} \label{figsc}
\end{figure}

Here we are going to analyse the tunneling conductance
$G=\frac{dI}{dV}$ as measured by STM.
In the absence of impurities, the contribution to the conductance
comes from the first term of Eq.\ \ref{cur1}. For s-wave
superconducting tips, one finds that the tunneling conductance
($G(V)=d{\mathcal I}/dV$) for $E_F>0$ and at $T=0$ is given by (with
$r=E_F/\Delta_0$, $p=-eV/\Delta_0$)
\begin{eqnarray}
G &=& G_0 \Big[{\mathcal N}_t(p)|r| + \int_{p} {\rm Sgn} (z-p+r)
{\mathcal N}_t(z)dz \Big] \label{freegra1} \\
\frac{dG}{dV} &=& \frac{eG_0}{\Delta_0} \Big[{\mathcal N}_t(p)
-{\mathcal N}_t^{'}(p)|r| - 2 \theta(p-r) {\mathcal N}_t(p-r) \Big] \nonumber\\
\label{freegra2}
\end{eqnarray}
where $ G_0= 8\pi^2e^2|U^{0}|^2(1+\xi^2) \rho_{0t} \rho_0/h$,
$\rho_G = \rho_0 |r-p|$, $\rho_t(r) =\rho_{0t} {\mathcal N}_t(r)$,
${\mathcal N}_t (x)= |x|/\sqrt{x^2-1} \theta(|x|-1)$, and ${\rm
Sgn}(x)$ denote the signum function. For graphene with $E_F=r=0$,
$dG/dV \sim {\rm Sgn}(V) {\mathcal N}_t(-V)$, {\it i.e.}, the tip
DOS is given by the derivative of the tunneling conductance. For
large $E_F$ away from the Dirac point, the first term of $G$ becomes
large and reflects the tip DOS. In between these extremes, when $E_F
\sim eV$, neither $G$ nor $dG/dV$ reflects the DOS. In this region,
the signature of the Dirac point appears through a cusp
(discontinuity) in $G\, (dG/dV)$ at $eV=-E_F-\Delta_0$ arising from
the contribution of the second (third) term in Eq.\ \ref{freegra1}
(Eq.\ \ref{freegra2}). These features, shown in Fig.\ \ref{figsc},
distinguishes such graphene STM spectra with their conventional
counterparts \cite{stmsc}.

\begin{figure}
\centerline{\psfig{file=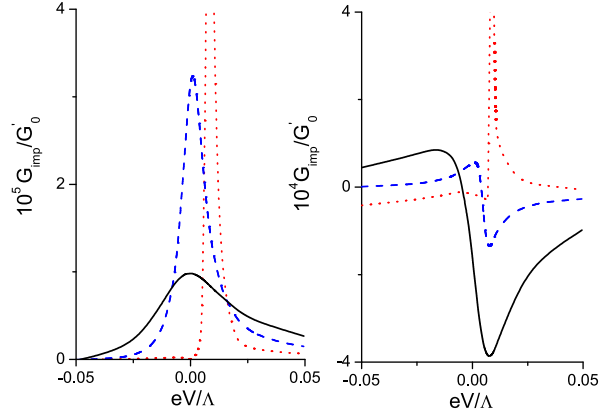,width=8cm}}
\caption{(Color online) Plot of $G_{\rm imp}$ as a function of $V$
for $|W^{0}/U^0|=0.05$ (right; impurity atop a site) and $2$ (left;
impurity atop hexagon center) for $E_F/\Lambda =0.3, 0.1, \,{\rm
and}\,0$ (black solid, blue dashed and red dotted lines
respectively). Plot parameters are $5U= V^{0}=0.05 \Lambda$, $W^0=
0.0005 \Lambda$, and $\epsilon_d =0$.} \label{fig3}
\end{figure}

Next, we turn to the case of impurity doped graphene and consider a
metallic tip with constant DOS. The contribution to the tunneling
conductance from the impurity (after subtracting the graphene
background) at $T=0$ (Eq.\ \ref{i1i2}) is
\begin{eqnarray}
G_{\rm imp} &=& G_0^{'}\frac{|B(V)|^2}{{\rm Im}\Sigma_d(V)}\frac
{|q(V)|^2-1+2 {\rm Re} [q(V)]\chi(V)}{\Lambda[1+\chi^2(V)]},
\label{imptun}
\end{eqnarray}
where $G^{'}_0= 2e^2 \rho_{0t}\Lambda/h$. Such tunneling
conductances are known to have peak/antiresonace/dip feature at zero
bias for $|q| \gg 1/\simeq 1/\ll 1$ \cite{madhavan1}. In
conventional metals, Eq.\ \ref{cur1} can be used to compute the STM
current by taking $U^0$ as a fixed parameter independent of the
position of the impurity. However, the situation in graphene
necessitates a closer attention to $U^0$ which is proportional to
the probability amplitude of the Dirac quasiparticles in graphene to
hop to the tip. The strength of $U^{0}$ can be estimated using the
well-known Bardeen tunneling formula \cite {bardeen1}: 
\begin{equation}
U^{0} \sim
\int d^2 r \left(\phi_{\nu}^{\dagger}(z)
\partial_z \Psi_G({\vec r},z) -\Psi_G^{
\dagger}({\vec r},z) \partial_z \phi_{\nu}(z)\right) \sim
\Psi_G({\vec r}_0,z_0)
\end{equation}
where the last similarity is obtained by a
careful evaluation of the surface integral $\int d^2 r$ over a
surface between the graphene and the tip parallel to the graphene
sheet \cite{tersoff1}, $({\vec r}_0,z_0)$ is the coordinate of the
tip center \cite{tersoff1}, $\phi_{\nu}(z)$ is tip electron
wavefunction, and the wavefunction graphene electrons $ \Psi_G
({\vec r},z)$ around $K (K')$ valley, can be written, within
tight-binding approximation, as \cite{mont1}
\begin{eqnarray}
\Psi_G ({\vec r},z) &=& \frac{1}{\sqrt{N}} \sum_{R_i^A} e^{ i
[\{\vec K (\vec K')+\vec {\delta k}\} \cdot {\vec R_i^A}]}
\Big[\varphi({\vec r}-{\vec R_i^A})
\nonumber\\
&& + e^{+(-)i \theta_k} \varphi({\vec r}-{\vec R_i^B}) \Big ] f(z).
\label{tightbinding}
\end{eqnarray}
Here $\theta_k = \arctan(k_y/k_x)$, $\vec {\delta k}$ is the Fermi
wave-vector as measured from the Dirac points with $|\vec {\delta
k}| \ll |\vec K( \vec K')|$ for all $E_F$, $\varphi({\vec r})$ are
localized $p_z$ orbital wavefunctions, $N$ is a normalization
constant, $f(z)$ is a decaying function of $z$ with decay length set
by work function of graphene, and $ R_i^{A(B)} = n {\hat a}_1 + m
{\hat a}_2 ({\hat a}_2 - \hat y)$ with integers $n,m$ denote
coordinates of the graphene lattice sites (Fig.\
\ref{ra_fig1})\cite{mont1}. When the impurity and the STM tip is atop
the center of the hexagon, pseudospin symmetry necessitates
$\varphi({\vec r}_0-{\vec R_i^{A,B}})$ to be identical for all
neighboring $A$ and $B$ sublattice points $1..6$ surrounding the
impurity (Fig.\ \ref{ra_fig1}). Consequently, the sum over lattice
vectors $R_i^A$ in Eq.\ \ref{tightbinding} reduces to a sum over the
phase factors $\exp (i [\{\vec K (\vec K')+ \vec {\delta k}\} \cdot
{\vec R_i^A}])$ for these lattice points. It is easy to check that
this sum vanishes for both Dirac points (when $|\vec {\delta
k}|=0$). Thus the only contribution to $\Psi_G({\vec r}_0,z_0)$
comes from the second and further neighbor sites for which the
amplitude of localized wavefunctions $\varphi({\vec r}_0-{\vec
R_i^{A/B}})$ are small. For finite $E_F$, ($\vec {\delta k} \ne 0$)
there is a finite but small contribution (${\rm O} (|\vec {\delta
k}|/|{\vec K|})$) to $\Psi_G({\vec r}_0,z_0)$ from the nearest
neighbor sites. Thus $\Psi_G({\vec r}_0,z_0)$ and hence $U^0$ is
drastically reduced when the impurity is atop the hexagon center. In
this case, we expect $U^{0} \ll W^0$ and hence $|q| \gg 1$ (Eq.\
\ref{qeq}) leading to a peaked spectra for all $E_F$. In contrast,
for the impurity atom atop a site, there is no such symmetry induced
cancellation and $\psi_G({\vec r}_0,z_0)$ receives maximal
contribution from the nearest graphene site directly below the tip.
Thus we expect $|U^0| \gg |W^0|$ (since it is easier for the tip
electrons to tunnel to delocalized graphene band than to a localized
impurity level) leading to $q \simeq I_1/I_2 \simeq
-\ln|1-\Lambda^2/(eV+E_F)^2|/\pi$. For large $|eV+E_F|$ and impurity
atop a site, $q \le 1$ leading to a dip or an antiresonance in
$G_{\rm imp}$ which is qualitatively distinct from the peaked
spectra for impurity atop the hexagon center. As $E_F \to 0$, $q$
diverges logarithmically for small $eV$. However, it can be shown
that in this regime $\chi$ shows a stronger linear divergence for
$eV \ne \epsilon_d$ which suppresses $G_{\rm imp}$. At
$eV=\epsilon_d$, the divergence of $\chi$ also becomes logarithmic
and we expect a peak of $G_{\rm imp}$. Note that these effects are
independent of $\Sigma_d$ and hence of the precise nature of the
impurity. Such an impurity position dependent peak/dip structure of
$G_{\rm imp}$ has been observed for magnetic impurities in Ref.\
\onlinecite{hari1} for $E_F \gg eV$.

To demonstrate this feature, we restrict ourselves to impurities
with small Hubbard $U$ and compute the self energy of the impurity
electrons within a mean-field theory where $U n_{\sigma} n_{\bar
\sigma} = U \langle n_{\sigma} \rangle n_{\bar \sigma}$ leading to
spin-dependent on-site impurity energy $\epsilon_{\sigma} =
\epsilon_d +U \langle n_{\bar \sigma} \rangle$ \cite{neto2}. Using
Eqs.\ \ref{hamil1} and \ref{graimp}, one then obtains the mean-field
advanced impurity Green function ${\mathcal G}^{{\rm
imp}}_{\sigma}(\omega) = (\omega -
\epsilon_{\sigma}-\Sigma_d(\omega))^{-1}$ where the impurity
self-energy is given by $\Sigma_d(\omega) = |V^0|^2 (I_1+i I_2)$ and
mean-field self-consistency condition demands $n_{\sigma} = \int
d\omega/\pi {\rm Im} {\mathcal G}^{{\rm imp}}_{\sigma}(\omega)$.
Following Ref.\ \onlinecite{neto2}, we solve these equations to get
$\chi(\epsilon)$, and ${\rm Im} \Sigma_d(\epsilon)$ which can be
substituted in Eq.\ \ref{imptun} to obtain $G_{\rm imp}$. We note,
from Eqs.\ \ref{imptun} and \ref{qeq}, that $G_{\rm imp}/G_0^{'}$
depends on the ratios $E_F/\Lambda$, $V^0/\Lambda$, and $W^0/U^0$
which can not be quantitatively determined from the Dirac-Anderson
model. We therefore treat them as parameters of the theory and
compute $G_{\rm imp}$ for their representative values as shown in
Fig.\ \ref{fig3}. In accordance with earlier discussions, we find
that for large $E_F/\Lambda=0.3$, $G_{\rm imp}$ has qualitatively
different features; for the impurity at the center of the hexagon,
it shows a peak (left panel) while for that atop a site (right
panel), it shows a dip. The change of $G_{\rm imp}$ from a dip to a
peak via an antiresonance as a function of $E_F/\Lambda$ when the
impurity is atop a site can be seen from right panel of Fig.\
\ref{fig3}. In contrast, the left panel always shows peak spectra.

Before ending this section, we note that the logarithmic divergence 
of the $G_{imp}$ when the impurity is reasonably close to the Dirac 
point is a characteristics of the Dirac physics of the low-energy 
quasiparticles. This feature is therefore also expected to be seen 
for tunneling conductance measured atop an impurity on the surface 
of a topological insulator. 

\section{Conclusion} \label{conclusion}
In this review we have presented a theory for transport properties 
across superconducting junctions of graphene with barriers of
thickness $d_0$ and arbitrary gate voltages $V_0$ applied across
the barrier region. 
The oscillatory behaviour of the tunneling conductance as well as 
Josephson current are shown to be robust even for a barrier of finite
width. In the thin barrier limit, such  behavior is the
manifestation of the transmission resonance of DBdG quasiparticles
in superconducting graphene. 
Graphene is an interesting candidate for transport applications, 
in particular for spintronics as it exhibits 
exhibits remarkably high mobility with easily controllable carrier density.
Superconducting junctions of graphene has recently been realized
experimentally \cite{heersche1,ref9}. Further experiments in this 
direction may therefore lead to realization of SBS and NBS junctions 
discussed in this review and hence may lead to verification of some the
theoretical results discussed here.

The effect of localized impurities on the electronic properties of graphene
has attracted a lot of recent attention. We have studied the effect of  
presence of localized magnetic impurities in graphene 
which gives rise to the Kondo effect i.e. the dynamic 
screening of the localized moment. The Kondo effect in graphene
is unconventional as the effective coupling Kondo coupling strength 
(for weak coupling regime) can be tuned by gate voltage. 
Recent studies \cite{uchoa3} have found that the Kondo coupling strength can 
also be controlled by a gate voltage in the strong coupling regime. 
We also discuss scanning tunneling conductance spectra
phenomenon for both doped and undoped graphene. The position of the 
impurity on or in graphene plays a subtle role and affects the
underlying physics of STM spectra in doped graphene. 
For impurity atoms atop the hexagon
center, the zero-bias tunneling conductance shows a peak; for
those atop a graphene site, it shows a dip. This feature is 
a direct consequence of pseudospin symmetry and Dirac nature of graphene
quasiparticles. 
A recent scanning tunneling microscopy (STM) experiment \cite{brar} has demonstrated the 
ability to controllably ionize individual Co adatom on graphene using either 
a back gate voltage or the STM tip bias voltage.
This has opened up the possibility of probing the interesting 
electronic phenomena which arises due to the interplay of the impurity with the graphene electrons.

We thank I. Paul and S. Bhattacharjee for collaboration on related topics. 
K.S thanks DST for support through grant SR/S2/CMP-001/2009.


\begin{thebibliography}{99}
\bibitem{comment1} Since our solution for the wavefunction at the
barrier region breaks down at $V_0=E_F$, it is necessary to
carefully analyze the limit $V_0 \to E_F$ for arriving at this
conclusion.
\bibitem{comment2} Note that at $V_0=E_F$,
$\alpha_c=0$ leading to $G=0$.
Thus the divergence of periodicity do not constitute a physical
singularity.
\end{thebibliography}
\end{document}